\documentclass[sigconf, screen, review]{acmart}

\AtBeginDocument{%
  }

 \setcopyright{none} 

\acmISBN{978-X-XXXX-XXXX-X/XX/XX}

\usepackage{mathtools}   

\usepackage{amssymb}

\usepackage{hyperref}

\usepackage{booktabs}
\usepackage[table]{xcolor}
\usepackage{array}

\usepackage{nicefrac}
\usepackage{siunitx}
\usepackage{array,framed}
\usepackage{booktabs}
\usepackage{
  color,
  float,
  epsfig,
  wrapfig,
  graphics,
  graphicx,
  subcaption
}
\usepackage{textcomp}
\usepackage{setspace}
\usepackage{latexsym,fancyhdr,url}
\usepackage{enumerate}
\usepackage{enumitem}
\setlist[itemize]{leftmargin=10pt}
\usepackage[ruled,vlined]{algorithm2e}
\usepackage{xcolor}
\usepackage{mdframed}

\usepackage{xcolor}
\newmdenv[
  backgroundcolor=green!10,
  linecolor=green!40!black,
  leftmargin=0,rightmargin=0,
  innerleftmargin=4pt,innerrightmargin=4pt,
  skipabove=4pt,skipbelow=4pt,
  roundcorner=2pt,linewidth=0pt
]{challengerbox}
\newmdenv[
  backgroundcolor=red!10,
  linecolor=red!40!black,
  leftmargin=0,rightmargin=0,
  innerleftmargin=4pt,innerrightmargin=4pt,
  skipabove=4pt,skipbelow=4pt,
  roundcorner=2pt,linewidth=0pt
]{workerbox}

\DontPrintSemicolon
\LinesNotNumbered 
\usepackage[most]{tcolorbox}

\usepackage[noend]{algpseudocode}

\newcommand*\circled[1]{\tikz[baseline=(char.base)]{
            \node[shape=circle,fill,inner sep=1pt] (char) {\textcolor{white}{#1}};}}
\usepackage{xcolor}
\usepackage{pgfplots}
\usepackage{tikz}
\pgfplotsset{compat=1.8}
\pgfplotsset{
    width=\textwidth,
}
\usepackage{lipsum}
\usepackage{pgfplotstable}
\usetikzlibrary{pgfplots.groupplots}

\usepackage{mathtools}

\DeclareMathAlphabet{\mathcal}{OMS}{cmsy}{m}{n}

\usepackage{svg}

\DeclareGraphicsExtensions{%
    .png,.PNG,%
    .pdf,.PDF,%
    .jpg,.mps,.jpeg,.jbig2,.jb2,.JPG,.JPEG,.JBIG2,.JB2}

\definecolor{byzantium}{rgb}{0.44, 0.16, 0.39}

\newif\ifshowcomments
\showcommentstrue 

\begin{document}

\title{Timing and Memory Telemetry on GPUs for AI Governance}
\author{Saleh K. Monfared}
\affiliation{%
  \institution{\small Worcester Polytechnic Institute}
  \country{USA}
}

\author{Fatemeh Ganji}
\affiliation{%
  \institution{\small Worcester Polytechnic Institute}
  \country{USA}
}

\author{Daniel Holcomb}
\affiliation{
  \institution{\small University of Massachusetts Amherst, USA}
}

\author{Shahin Tajik}
\affiliation{%
  \institution{\small Worcester Polytechnic Institute}
  \country{USA}
}

\begin{abstract}
The rapid expansion of GPU-accelerated computing has enabled major advances in large-scale artificial intelligence (AI), while simultaneously heightening concerns about the ability to observe, characterize, or govern how powerful accelerators are used once deployed.
Governance is essential to ensure that large-scale compute infrastructure is not silently repurposed for training models, circumventing usage policies, or operating outside legal or institutional oversight.
Because current GPUs expose limited trusted telemetry and can be modified or virtualized by resourceful adversaries, we explore whether compute-based measurements can provide actionable signals of utilization in settings where both the host and device may be untrusted.
We introduce a measurement framework that leverages architectural characteristics of modern GPUs to generate timing‑ and memory‑based observables that correlate with real compute activity.
Our design draws on four complementary primitives: (1) a probabilistic, workload-driven mechanism inspired by Proof‑of‑Work (PoW) to expose parallel effort, (2) sequential, latency‑sensitive workloads derived via Verifiable Delay Functions (VDFs) to characterize scalar execution pressure, (3) General Matrix Multiplication-based (GEMM)-based tensor‑core measurements that reflect dense linear‑algebra throughput, and (4) a VRAM‑residency test that distinguishes on‑device memory locality from off‑chip access through bandwidth-dependent hashing. 
These primitives offer statistical and behavioral indicators of GPU engagement that remain observable even without trusted firmware, enclaves, or vendor‑controlled counters. 
We evaluate their responses to contention, architectural alignment, memory pressure, and power overhead, demonstrating that timing shifts and residency latencies can reveal meaningful patterns of utilization.
Our results illustrate why compute-based telemetry can complement future accountability mechanisms by exposing architectural signals relevant to post-deployment GPU governance. 
\end{abstract}


\maketitle
\thispagestyle{plain} 
\pagestyle{plain}

\section{Introduction}\label{sec:intro}

The growth of GPU-accelerated computing has enabled advances in artificial intelligence (AI), powering developments in large language models, generative design, autonomous systems, and scientific discovery. 
However, this surge in GPU deployment has raised concerns about the governance of AI~\cite{dafoe2018ai,CAIS_AIStatement_2025}. 
As the risk of misuse or unintentional release of harmful capabilities continues to grow~\cite{urbina2022dual,bucknall2022current}, many approaches across technical, institutional, and policy dimensions are being explored to ensure the development of governable AI systems~\cite{bullock2024oxford,batool2025ai}.
Among these, hardware-enabled governance has emerged as a promising direction, leveraging the physical and architectural properties of computing hardware to support oversight and transparency~\cite{aarne2024secure,kulp2024hardware,sastry2024computing,petrie2025flexible,petrie2025technical,kembery2024towards,scher2025mechanisms}.
Although significant progress has been made, substantial technical challenges persist in deploying such mechanisms at scale~\cite{reuel2024open,Bengio_FlexHEG_Memo_2024}.
These include heterogeneous accelerator support, the absence of standardized measurement interfaces, and the difficulty of integrating hardware trust anchors into existing AI infrastructure~\cite{barnett2025technical,petrie2025technical}.

Current commercial GPUs lack tamper-evident or trustworthy mechanisms for usage monitoring.
Even high-end accelerators such as the NVIDIA H100 offer no built-in capabilities for sealed logging or hardware-level oversight.
Firmware stacks remain vendor-signed and closed-source, on-device counters can be reset, forged, or suppressed by a privileged host, and no native features exist for privacy-preserving rate limiting or post-deployment accountability.
Once a GPU is shipped or integrated into a distributed or decentralized compute infrastructure, the vendor loses visibility into whether the hardware is used in accordance with declared intent.

Embedding telemetry within Trusted Execution Environments (TEEs) initially appears attractive, as enclaves could in principle protect measurement logic from a compromised host.
However, TEEs such as Intel SGX and AMD SEV have been repeatedly broken by practical microarchitectural attacks (e.g., Foreshadow~\cite{van2018foreshadow}, CacheOut~\cite{van2021cacheout}) and fault-based exploits (e.g., Plundervolt~\cite{murdock2020plundervolt}, SEV-Voltage Glitch~\cite{buhren2021one}).
These attacks highlight a persistent mismatch between TEE threat models and real-world adversaries.
Similarly, NVIDIA’s Confidential Computing (CC), e.g., H100 CC mode~\cite{nvidia_HCC_whitepaper_2023}, does not fully address well-resourced attackers.
Recent work has reverse-engineered CC internals~\cite{gu2025nvidia} and demonstrated practical breakages~\cite{chuang2025teefail} at low cost.
Nvidia explicitly mentioned these attacks as out of scope: Advanced physical and side-channel attacks~\cite{kiyan2024through,krachenfels2021automatic,mehta2025swarm}, limiting the robustness of CC-based solutions. 

This lack of reliable visibility into utilization poses challenges for AI governance and safety.
In the absence of trusted hardware telemetry, powerful GPUs can be silently repurposed to train or execute AI models for malicious purposes, including coordinated cyberattacks, automated vulnerability exploitation~\cite{anthropic_AI_espionage_2025,singh2025artificial}, advanced malware generation~\cite{metta2024generative}. 
At the same time, utilization monitoring must preserve user privacy and avoid leaking sensitive model architectures, training data, or application logic.

\begin{figure}[t]
\centering
\includegraphics[width=0.8\columnwidth]{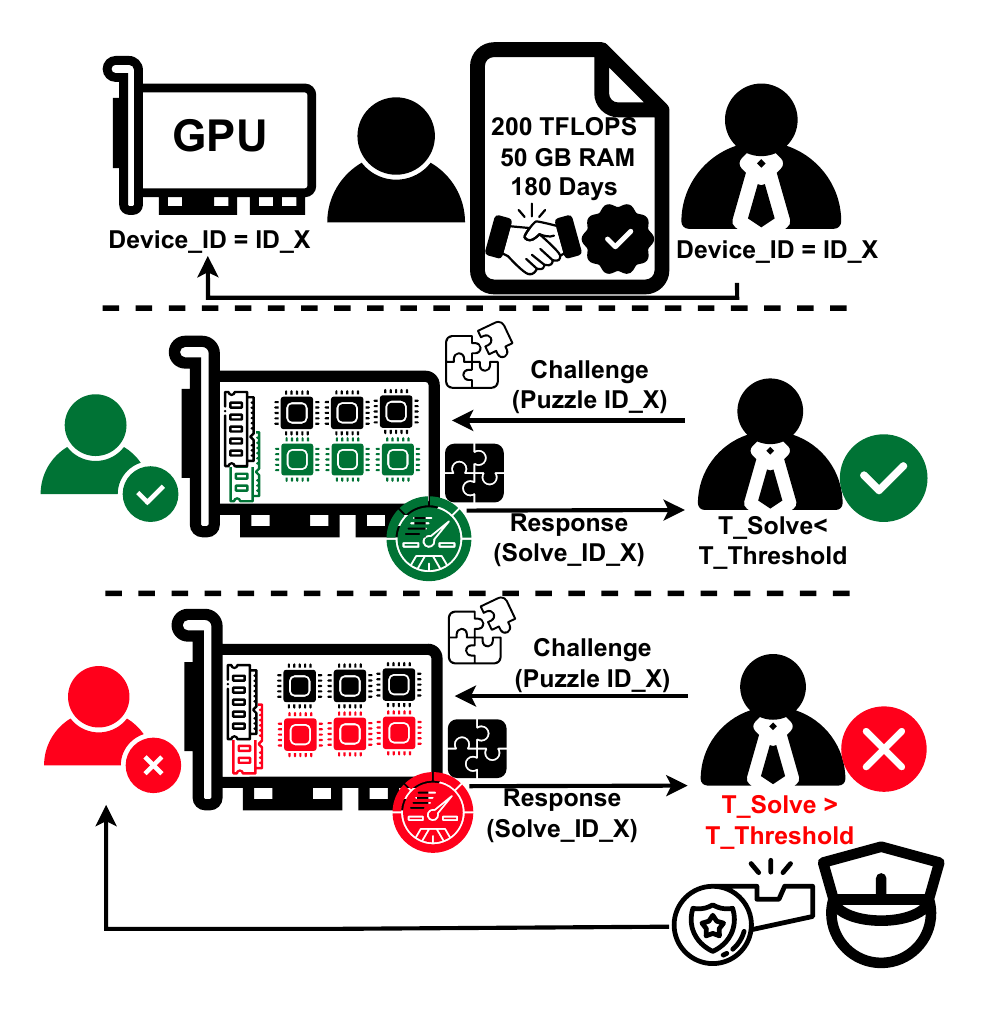}
\caption{Time- and memory-intensive workloads can induce measurable utilization effects without relying on trusted hardware.} 
\label{fig:high_level}
\end{figure}

In light of these constraints, we ask a more general question: 
\textit{can GPU utilization be inferred through measurable computational effort and memory behavior, even in adversarial environments without trusted telemetry?} 
Rather than relying on hardware trust anchors, we explore whether time-, memory-, and architecture-dependent measurements can provide practical indicators of real activity; see Figure~\ref{fig:high_level}. 
Reiterating the limitations of hardware-based enforcement, we shift focus from secure monitoring to compute-based measurement.
In the most general setting where neither the host nor the GPU can be trusted, we explore four measurement primitives for \textit{remote utilization inference}.
These primitives draw on Verifiable Delay Functions (VDFs)~\cite{boneh2018verifiable} to induce sequential latency, Memory-Hard Functions (MHFs)~\cite{biryukov2016egalitarian} to stress HBM and compute pathways, and dense linear algebra workloads, such as large-scale General Matrix Multiplication (GEMM), to activate tensor cores and create measurable load patterns.
The goal is not to design a protocol or enforce compliance, but to evaluate whether such workloads generate behavioral signatures that correlate with actual GPU utilization.

\noindent\textbf{Contributions.} 
We formalize the problem of \textit{compute activity inference} during post-deployment in adversarial settings, and describe why existing hardware-security approaches (e.g., NVIDIA CC) remain insufficient for visibility against highly resourced adversaries. 
In doing so, the following contributions are made. 
\begin{enumerate}
\item We adapt well‑known mechanisms, such as Proof of Work (PoW) and memory-hard functions, to GPU‑specific architectural behaviors, e.g., HBM load, tensor‑core saturation, etc., thereby binding classical theory to concrete, architecture-dependent observables that prior work has not exploited. 
We design and evaluate four compute-based measurement primitives, namely sequential workloads, memory-hard functions, GEMM-based tensor utilization, and VRAM residency measurements, which induce architecture-dependent behavior correlated with real GPU activity. 
\item We analyze these primitives with respect to timing latency, power footprint, memory usage, and scheduling interference, quantifying how observable signals respond to co-residency with realistic workloads, such as LLM inference and remote execution.
\item We demonstrate the viability of these measurements in constrained, noisy, and virtualized settings, establishing that even without hardware trust anchors, passive telemetry can support future governance use cases. 
\end{enumerate} 


\section{GPU in AI Workloads}\label{sec:backgraound}

Originally designed for graphics processing, modern GPUs have evolved into massively parallel architectures optimized for matrix-heavy operations in deep learning and scientific computing~\cite{sze2017efficient,reuther2020survey}.
Unlike CPUs with a few general-purpose cores, GPUs use a Single Instruction, Multiple Threads (SIMT) model to execute arithmetic operations concurrently across large datasets~\cite{lindholm2008nvidia,nvidia2011nvidia}.
This enables high-throughput performance for linear algebra workloads, including convolutional neural networks, transformers, and generative diffusion models.

Data center GPUs, such as NVIDIA's Ampere and Hopper~\cite{NVIDIA_H100_2022,cui2025story} and AMD's Instinct CDNA-based accelerators, combine tensor units, thousands of compute lanes, and high-bandwidth memory to accelerate mixed- and reduced-precision arithmetic (e.g., FP16, BF16, FP8)\cite{NVIDIA_T4_datacenter}.
Deployed at scale using interconnects like NVLink and NVSwitch, they support data- and model-parallel execution\cite{sergeev2018horovod}.
System efficiency depends on more than raw compute; it hinges on memory bandwidth, interconnect topology, and workload scheduling~\cite{li2019evaluating}.
As a result, GPU utilization now serves as a key indicator of performance, energy cost, and governance relevance, especially in unregulated settings where misuse is possible.

GPU virtualization is central to multi-tenant AI infrastructure.
Mechanisms like Multi-Instance GPU (MIG)\cite{NVIDIA_MIG}, Multi-Process Service (MPS)\cite{NVIDIA_MPS_doc}, and SR-IOV~\cite{single2010single} allow partitioning across users, containers, or VMs, while attestation tools, TPMs~\cite{kinney2006trusted}, Intel TDX~\cite{cheng2024intel,Intel_TDX_overview}, AMD SEV-SNP~\cite{AMD_SEV_overview}, and NVIDIA CC~\cite{nvidia_HCC_whitepaper_2023}, verify workload provenance and isolation.
Yet these approaches do not reveal how resources are actually used.
Virtualization abstracts low-level telemetry and introduces side-channels, residual state, and timing leakage~\cite{dutta2023spy,zhang2024beyond,zhang2025nvbleed}, preventing attribution of compute usage or detection of covert repurposing.

This motivates \emph{telemetry-based utilization measurement}, not as a substitute for  attestation, but as a complementary signal for enforcement, anomaly detection, or fairness auditing.
These mechanisms do not intend to provide cryptographic guarantees; instead, they expose observables such as compute rate, memory bandwidth, or solve-time distributions.
Our measurements rely on no hardware trust anchors, but expose timing and memory patterns for future governance protocols.

\section{Untrusted GPU Execution Settings}
\label{sec:threat_model}

We consider a range of deployment cases in which external entities may wish to observe or estimate GPU usage for purposes such as telemetry logging, policy enforcement, or future integration into accountability protocols. 
The goal is to examine whether puzzle-based measurements (e.g., solve times, memory access penalties) provide a useful signal under minimal trust assumptions. 
The measurements we study are intended as heuristic indicators of usage, rather than full adversarial resistance or proofs.  
To this end, we define four representative settings that differ in which parts of the software/hardware stack can be assumed trustworthy. 

\textbf{Case~{1}: Untrusted host and untrusted GPU.}
This is the most adversarial and general configuration. 
Both the host and GPU firmware may be modified, virtualized, or colluding, leaving no guarantees of trusted telemetry or execution. In this setting, traditional remote attestation fails, but puzzle-based measurements may still detect resource contention or cold-access penalties through statistical timing effects.
\textbf{Case~{2}: Host-side TEE only.}
The host runs within a Trusted Execution Environment (TEE), whereas the GPU remains untrusted. 
This model provides a narrow trust root for measurement orchestration (e.g., nonce generation, response logging), but cannot ensure correct GPU behavior. 
Authenticated metric protocols~\cite{ivanov2023sage,anderson1998new} may help record measurements, but the GPU remains an opaque device.
\textbf{Case~{3}: Confidential computing mode.}
Both host and GPU participate in a vendor-supported secure channel (e.g., NVIDIA CC~\cite{nvidia_HCC_whitepaper_2023}). While this provides workload confidentiality and basic telemetry sealing, firmware-level control and resource accounting are not guaranteed, and puzzle-based measurements may help fill observability gaps.
\textbf{Case~{4}: Trusted host, untrusted GPU.}
In edge-compute or sovereign-control deployments, the host may be secure (e.g., tamper-resistant or boot-locked), but the attached GPU is untrusted. Measurement primitives issued by the host may still yield coarse-grained telemetry if the GPU exhibits detectable contention or memory-access timing variations.
Although cases~{2}–4 offer varying degrees of baseline trust, they often lack visibility into fine-grained GPU usage. 
Case~{1}, being the most adversarial, motivates our focus: can simple measurements, without hardware roots of trust, still expose useful signals for future enforcement or logging? 
This paper explores that question from a measurement-design standpoint. 

\section{Timing Measurements}\label{sec:approach}

This section introduces workload-based timing measurements that function as telemetry probes, providing soft-enforcement signals in untrusted environments without TPMs, TEEs, or trusted telemetry.   
These mechanisms aim to detect whether a worker (e.g., a GPU) has executed a specified unit of computation, when it occurred, and under what degree of contention.  
These measurements enable latency-based inference, which may inform governance-relevant mechanisms, such as job-scheduling integrity or co-location detection.  
Each workload probe reveals a different dimension of device behavior: tensor-core saturation (GEMM), parallel effort (PoW), or sequential pressure (VDF).  
A randomized challenge schedule enables these probes to be sampled over time, providing the challenger with temporal visibility into resource engagement.  

\subsection{Latency Model}
This subsection establishes the core timing architecture used by the challenger to infer activity from the worker.  
The unified latency model defines how challenges are issued, how solve times are interpreted, and how randomness is injected into the measurement process.  
Presenting this framework upfront contextualizes the use of PoW, VDF, and GEMM probes as latency-based observables that underpin subsequent sections.  

We employ compute- and memory-intensive primitives that are infeasible to solve under realistic adversarial constraints faster than within a bounded time window.  
When coupled with explicit timing constraints, these primitives enable the challenger to infer that a given task was executed within a specific period.  
This establishes measurement reliability by relying on assumptions about effort and sequentiality~\cite{goldreich2005foundations,Pricing,back2002hashcash}, rather than on conventional cryptographic properties such as confidentiality or authenticity.  
Unlike trusted telemetry or secure co-processors, often susceptible to virtualization, spoofing, or insider threats, our model relies on the statistical detectability of timing deviations.  

To support this model, we adopt challenge–response models in which the challenger issues compute tasks and monitors solve latency as the primary observable.  
These tasks are drawn from a structured workload family varying in parallelism, sequential depth, and architectural alignment: PoW challenges stress hash pipelines, VDFs measure sequential compute capacity, and GEMM kernels exercise tensor-core throughput. 
We emphasize that we draw on these established workload patterns as inspiration for constructing timing probes. 
They can be replaced by their lightweight peers that serve the same purpose, thereby reducing the computational and energy costs of existing mechanisms. 
The following subsections detail the design and role of each workload type, PoW, VDF, and GEMM, within this timing-based telemetry framework.

\noindent\textbf{Tunable Parallel Load.}
Here we focus on a \emph{PoW-style} mechanism to measure GPU utilization where the challenger periodically issues computation puzzles to the worker (the GPU) and checks both the \emph{correctness} and the \emph{latency} of the returned solutions. 
PoW-style constructions have a long history outside cryptocurrencies, for example, they are used to rate-limit abuse and mitigate automated clients in web services (e.g., anti-bot measures on high-value web endpoints~\cite{gulihar2018anomaly,kaiser2008mod}). 

A PoW-style puzzle requires the worker to find an input $x$ such that the output of a cryptographic hash function $H(x)$ satisfies a target condition
$H(x) < \text{target}$, where the target defines the puzzle's difficulty. 
The challenger can efficiently confirm the solution by recomputing $H(x)$, while the worker must perform a brute-force search over many candidate inputs $\{x_i\}$ until the inequality holds, making the expected computational effort proportional to the inverse success probability \(1/p = 1/\Pr[H(x) < \text{target}]\). 
By tuning puzzle difficulty, the challenger can probabilistically enforce an expected share of device resources. 
Difficult puzzles require more sustained parallel work and therefore imply greater occupation of GPU cores and memory during the puzzle interval. 

Our approach leverages a measurement scheme rooted in memory-hard hash functions. 
We model PoW solve times under both single-thread and multi-thread (GPU) settings, showing the exponential-time behavior and memoryless property that enables a probabilistic mechanism. 
This method enforces a relationship between GPU utilization and the computational resources consumed during execution. 
By employing memory-hard primitives such as Argon2\cite{biryukov2016argon2}, the scheme ensures that a significant portion of the high-bandwidth memory (HBM) is actively engaged during the measurement process. 
A practical measurement procedure follows a pattern similar to use cases in cryptocurrencies: the challenger samples a strong random salt from a TRNG, composes a puzzle input that includes the salt and a trusted timestamp, and transmits it to the worker. 
The worker must perform a large-scale, parallel search or time-bound computation over the input (using thousands of GPU threads) and return a solution within the prescribed time window. 
The challenger validates the returned solution for correctness and checks that the observed response time lies within the expected interval. 
Deviations indicate possible under-utilization, outsourcing, or misbehavior. 
Although this mechanism introduces considerable power and thermal overhead, it provides a quantifiable effort metric of sustained memory utilization within the target GPU, making it well-suited for scenarios with the need for checking resource allocation. 
Shifting from PoW to GEMM can reduce power consumption and thermal overhead by aligning computation with native tensor-core execution, which is more efficient. 

\noindent\textbf{Compute Saturation.}
Generic PoW constructions alone are insufficient to target tensor-core execution.
To avoid fallback to slower SIMT scalar units, used, we employ GEMM challenges tailored for native execution on Tensor Cores~\cite{komargodski2025proofs,braverman2025sublinear}. 
Modern frontier-scale AI models rely overwhelmingly on tensor-core execution rather than traditional scalar CUDA cores.
Prior architectural studies~\cite{nvidiaVolta,nvidiaAmpere, NVIDIA_H100_2022, geiping2023cramming} show that training and inference of transformers, LLMs, and diffusion models consist almost entirely of dense linear-algebra kernels, particularly GEMM and attention-projection matrix multiplications, whose throughput is dominated by specialized Tensor Core units (e.g., NVIDIA's Warp Matrix Multiply-Accumulate pipelines).
In fact, empirical workload analyses~\cite{du2025bitdecoding, anderson2023optical} demonstrate that over \textit{90\%} of floating-point operations in LLMs involve GEMM, which are efficiently executed on Tensor Cores using FP16, BF16, or FP8 pathways, with scalar CUDA cores contributing minimally to total FLOPs.

The GEMM uses Warp Matrix Multiply-Accumulate (WMMA) and Hardware Matrix Multiply-Accumulate (HMMA) instructions.
By analyzing both solve-time distribution and output correctness, the challenger can distinguish tensor-core execution from scalar emulation with high confidence.
This ensures that the workload engages the intended high-throughput matrix-multiply hardware.

The challenger samples two pseudorandom matrices derived from the salt and timestamp and instructs the worker to perform a sequence of dimension-controlled GEMM operations whose total FLOP count is tightly coupled to the requested utilization level.
Because GEMM saturates both compute units and high-bandwidth memory, the resulting latency is highly reproducible on a local GPU but difficult to outsource to CPUs, small GPUs, or remote clouds.
Moreover, the challenger can cheaply validate correctness by hashing the returned matrix or verifying random linear combinations of rows and columns.   
This GEMM-based PoW thus provides a practical, hardware-aligned mechanism for continuous utilization measurement that exploits the deterministic throughput characteristics of modern GPU matrix engines.

\noindent\textbf{Sequential Delay.} 
Shifting from PoW to Verifiable Delay Function (VDF)~\cite{boneh2018verifiable,wesolowski2020efficient,rivest1996time} enables the challenger to capture not only parallel effort but also sequential compute pressure, offering finer-grained insight into whether the worker sustained uninterrupted execution over time.
This capability is essential in governance scenarios where precomputation or parallel outsourcing must be disincentivized.
To this end, we deploy a multi-instance VDF strategy that measures the effective sequential compute capacity of GPUs.
Our design is optimized for GPU execution, minimizing kernel divergence and enhancing predictability.
In particular, we adopt a construction that balances simplicity and efficiency~\cite{wesolowski2020efficient}, which is well-suited for continuous utilization monitoring in a multi-instance setting across thousands of GPU cores.
Each VDF instance is solved independently and sequentially, ensuring that parallelism provides throughput gains only within controlled architectural limits.
At regular intervals, the challenger generates randomized VDF puzzles and transmits each input, including a securely generated salt and timestamp, to the worker.
The GPU then performs multi-instance VDF evaluations locally by executing sequential squaring operations over large modular arithmetic groups.
Upon completion, the worker returns both the computed outputs and associated proof elements to the challenger. 
The challenger then checks both the \textit{correctness} of the computation and the \textit{timing constraints} associated with the response.
To mitigate the risk of outsourcing to alternate platforms such as CPU clusters or ASICs, one tunes instance difficulty, modulus size, and iteration depth to favor GPU architectures.

\begin{algorithm}[t]
\small
\caption{Continuous Compute-Based Measurement}
\label{alg:cont}
\DontPrintSemicolon
\KwIn{
minimum target rate per round $\lambda_{\min}$;\;
instances for VDF Puzzles $t_{\text{vdf}}$;\;
threads per instance $T$;\;
total rounds $n$;\;
round interval $\Delta$;\;
latency offset $t_0$
}
\KwOut{Decision on $H_0:\ \lambda \ge \lambda_{\min}$}

$S \gets 0$  \tcp*{Sum of adjusted per-round solve times}
$R \gets 0$  \tcp*{Number of completed rounds}


  \begin{challengerbox}
    \textbf{Challenger.} Generate $t$ ($t_{\text{vdf}}$ or $t_{\text{pow/gemm}}=1$) salted challenges $\{\mathsf{ch}_j\}_{j=1}^{t}$ with TRNG salts.  
    Start timer $t_{\text{start}} \gets \mathsf{Now}()$.  
    Send all challenges in one batch to the worker.
  \end{challengerbox}

  \begin{workerbox}
    \textbf{Worker (GPU).} Launch a kernel with $t$ instances, each using $T$ threads.  
    For each instance $j$: compute output $\mathsf{sol}_j$.  
    Compute a single aggregate hash $\mathsf{H}_{\text{agg}} \gets H(\mathsf{sol}_1 \,\|\, \cdots \,\|\, \mathsf{sol}_t)$.  
    Return  $\mathsf{H}_{\text{agg}}$ to the challenger.
  \end{workerbox}

  \begin{challengerbox}
    \textbf{Challenger.} Receive $\mathsf{H}_{\text{agg}}$ and check validity (according to issued challenges).  
    $T_{\mathrm{round}} \gets \mathsf{Now}() - t_{\text{start}}$.  
    If $t_0>0$, set $T_{\mathrm{round}} \gets \max(T_{\mathrm{round}} - t_0, 0)$.  
    If valid: $S \gets S + T_{\mathrm{round}}$, $R \gets R + 1$.  
    Sleep until next round $(t_{\text{start}} + \Delta)$.
  \end{challengerbox}

\textbf{Decision.} Compute average round time $\bar{T} = S / R$.  
Set threshold $\tau = 1/\lambda_{\min}$ (maximum allowed average time per round).  
If $\bar{T} \le \tau$, \textbf{ACCEPT} $H_0$; else, \textbf{REJECT} $H_0$.
\end{algorithm}

\subsection{Continuous Compute-based Measurement} 
The challenger continuously sends the worker randomized, time-bound puzzles, enforces an expected utilization level by adjusting difficulty probabilistically, and resists trivial outsourcing by building puzzles that exploit GPU architectural advantages (massive parallelism, tensor cores, and high memory bandwidth) via memory-hard functions, GEMMs, and multi-instance VDFs. 
Verification checks both solution validity and timing to detect overuse or policy violations.

\begin{figure*}[t]
\centering
\includegraphics[width=0.85\textwidth]{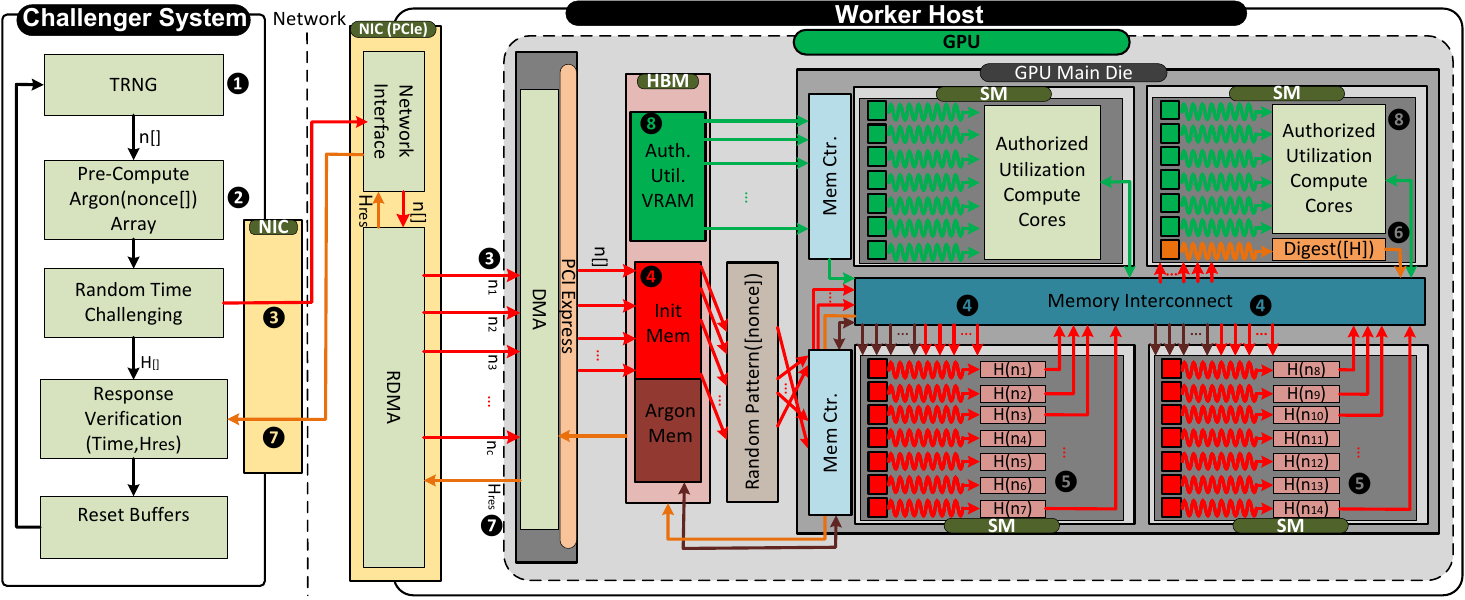}
\caption{High-level illustration of a randomized Challenge-Response mechanism that exploits HBM latency on the GPU. \vspace{10pt}}
\label{fig:overview}
\end{figure*}

Preventing outsourcing (offloading the puzzle to CPUs, remote clusters, or ASICs) is central to the construction. 
To raise the cost of outsourcing and to favor massively parallel GPU execution, puzzles combine (1) \emph{memory-hard} functions that require large working sets and high memory bandwidth (making ASIC/CPU implementations relatively inefficient), (2) high-throughput \emph{GEMM-based workloads} executed on tensor cores to exploit specialized GPU matrix units unavailable on commodity CPUs, and (3) \emph{multi-instance VDF/time-lock} constructs that enforce sequential or latency-bounded work across many parallel instances.


Algorithm~\ref{alg:cont} gives a high-level flow of the proposed continuous compute-based measurement procedure. 
Each round begins when the {challenger} generates salted challenges and sends them in a single batch to the {worker (GPU)}. 
The GPU solves the PoW/GEMM puzzle or executes multiple VDF instances in parallel, each launched with $T$ threads. 
After computation, the GPU returns a single aggregate hash (or GEMM output) to the challenger. 
The challenger validates this solution and records the total round latency to verify compliance with the expected utilization threshold. 
If the average round time $\bar{T}$ remains below the allowed bound $\tau = 1/\lambda_{\min}$, the measured utilization is accepted; otherwise, a deviation or policy violation is detected.

\subsection{Random-Timed Challenge-Response}

In contrast to the continuous approach, which incurs inherent unavoidable power and computation overhead, we propose a random-timed Challenge-Response mechanism. 
For this purpose, we focus on the architectural features that are particularly ly belonged to GPUs. 
Specifically we build upon our approach based on GPU's high-bandwidth and large memory. 
Furthermore, we emphasis the heavy memory transitions in today's LLMs on cutting edge GPUs and select the memory utilization for measurement.





Figure~\ref{fig:overview} shows a high-level flow of the proposed challenge–response mechanism, which explicitly exploits the properties of the GPU's high-bandwidth memory (HBM).  
In this scheme, a pre-determined \textit{challenge data block} is stored inside the GPU's VRAM before the measurement process begins.  
This part, highlighted in red in Figure~\ref{fig:overview} and referred to as the \textit{Init. Mem.}, serves as the trusted memory base for all subsequent computations.  
Any violation of this storage requirement (e.g., moving challenge data off HBM) invalidates the timing assumptions.  
Our detection method relies on the observation that the computation delay of a given hash function operating on data locally stored in HBM is statistically lower than when the same data must traverse off-chip over PCIe.  
This latency gap forms the basis for identifying compliant versus non-compliant behavior.

The overall framework operates as follows.  
In \circled{1}, a challenge generation loop is initiated by the challenger, using a true random number generator (TRNG) to create multiple nonce instances.  
These nonces are packed into an array as shown in \circled{2} and transmitted over the network to the worker.  
Upon reception, the packets arrive at a PCIe-enabled NIC and are passed directly via RDMA to the GPU's DMA engine, where they are placed into HBM (VRAM) with minimal latency in \circled{3}.  
Next, in \circled{4}, the pre-stored initial memory data (Init. Mem.) is accessed and used to seed the computation.  
In our implementation, we instantiate multiple concurrent instances ($C$) of the \texttt{Argon2id} hash function in keyed mode, where the initial memory data is mixed into each hash block (i.e., \texttt{BLAKE2b}) to prevent precomputation attacks.  
Following the contract and policy constraints, the worker GPU launches $T$ threads to compute all challenge instances in parallel as indicated in \circled{5}.  
Then, in \circled{6}, The worker computes memory-hard responses and returns the timing and hash result. 
Once all parallel computations are complete, the outputs are concatenated and digested using a fast aggregation hash (e.g., \texttt{SHA-256}), and the resulting single digest is sent back to the challenger through the network (\circled{7}).  
On the host side, the challenger checks both the \textit{correctness} of the digest and the \textit{timing consistency} with the expected bounds, then resets the state for the next challenge, which is issued at a random interval.  
During the challenge period, the GPU may continue to perform other computations or launch kernels concurrently (\circled{8}), ensuring that measurement minimally interferes with normal workload execution while still enabling query-based monitoring.

\subsection{Comparison of the proposed methods}

\begin{table*}[!t]
\caption{Overall comparison of the proposed measurement mechanisms.}
\label{instructions}
\begin{center}
\small
\renewcommand{\arraystretch}{1.3}
\setlength{\tabcolsep}{7pt}
\rowcolors{2}{gray!5}{white}
\resizebox{0.8\textwidth}{!}{%
\begin{tabular}{|c|c|c|c|c|}
\hline
\rowcolor{gray!20} 
\textbf{Metric / Approach} 
& \textbf{Proof-of-Work} 
& \textbf{GEMM Puzzle} 
& \textbf{Verifiable Delay Func.} 
& \textbf{HBM Challenge-Response} 
\\ \hline

\textbf{Measurement Type} 
& HBM/CUDA Compute 
& Tensor Compute  
& CUDA Compute 
& HBM 
\\ \hline

\textbf{Methodology} 
& \cellcolor{green!15}Complexity-enforced 
& \cellcolor{green!15}Complexity-enforced   
& \cellcolor{green!15}Complexity-enforced  
& \cellcolor{yellow!15}Architectural 
\\ \hline

\textbf{Power Overhead} 
& \cellcolor{red!15}Continuous Compute/Mem  
& \cellcolor{red!15}High Tensor-Core Activation 
& \cellcolor{yellow!15}Continuous Compute 
& \cellcolor{green!15}Random HBM Stress 
\\ \hline

\textbf{Detection Time} 
& \cellcolor{red!15}High (Multiple Samples) 
& \cellcolor{red!15}High (Multiple Samples) 
& \cellcolor{yellow!15}Medium (Predictable Delay) 
& \cellcolor{green!15}Low (Instant Detection) 
\\ \hline

\textbf{Outsource Complexity} 
& \cellcolor{green!15}High 
& \cellcolor{yellow!15}Medium 
& \cellcolor{yellow!15}Medium 
& \cellcolor{green!15}High 
\\ \hline
\end{tabular}%
}
\end{center}

\label{tab:gpu_attestation_summary}
\end{table*}

Table~\ref{tab:gpu_attestation_summary} summarizes the key distinctions between the proposed GPU measurement methods.
Each mechanism offers a different trade-off between detection latency and operational overhead. 
The PoW-based measurement relies on continuous compute-bound puzzles to verify GPU activity over time.  
It provides \textcolor{green!60!black}{complexity-enforced} high resistance to outsourcing, as solving the puzzles efficiently requires a large degree of parallelism.  
However, the method incurs a \textcolor{red!60!black}{high power overhead} due to sustained GPU load and exhibits slower detection since multiple samples are required to infer utilization trends.

In addition to hash-based PoW, {tensor-core GEMM puzzle} exploits the GPU's dense-linear-algebra units rather than {scalar ALUs in CUDA cores}.
Tensor cores deliver extremely high throughput for mixed-precision matrix multiplications, enabling puzzles with \textcolor{green!60!black}{computation/performance}, and very high \textcolor{green!60!black}{outsourcing resistance}, but it comes at consistently \textcolor{red!60!black}{high power overhead}.
Since CPUs, ASICs, and remote clusters cannot efficiently emulate tensor-core GEMM throughput, this puzzle naturally enforces device-local execution and enables reliable utilization measurement on modern GPUs.

On the other hand, the VDF approach ensures verifiable computation delay through sequential cryptographic operations, offering a balanced middle ground between PoW and CRP.  
It provides \textcolor{green!60!black}{complexity-enforced} detectability, similar to PoW, but with a more predictable timing model and a moderate power cost. 
Because computation proceeds sequentially across multiple instances, detection latency is lower than with PoW but still higher than HBM-based methods.

The CRP method leverages the architectural properties of the GPU's \textcolor{green!60!black}{High Bandwidth Memory (HBM)} rather than relying solely on computation-related assumptions.  
By verifying whether challenge data resides locally in HBM, it detects off-chip data accesses with minimal delay, achieving \textcolor{green!60!black}{low detection time} and negligible continuous power overhead.  
While it primarily offers \textcolor{yellow!60!black}{architectural} rather than compute-based enforcement, it provides strong resistance to outsourcing and enables near-instant anomaly detection via timing deviations between on-chip and PCIe-bound operations.




\section{Memory and Residency Inference}
\label{sec:residency_inference}

Residency inference complements timing-based telemetry by shifting the focus from how long to where a workload was executed, specifically, whether GPU-local memory (HBM) was used instead of host or remote memory.
This spatial distinction is vital for AI governance, revealing evasive behaviors such as memory outsourcing, virtualization, or interference that compromise isolation and traceability.
Combined with latency signals (Section~\ref{sec:approach}), residency enables a dual-axis telemetry framework capturing both temporal engagement and spatial locality. 

\noindent\textbf{Argon2id-based VRAM test.}
As an architectural solution to mitigate the power overhead in continuous measurement, we propose a \textit{residency-aware, bandwidth-bound measurement primitive} based on the keyed-mode \texttt{Argon2id} function.  
Our design relies on an \emph{incompressible challenge dataset} (\textsc{CHAL}) stored in GPU VRAM and processed by a cryptographic hash function that requires randomized access to the dataset.
This ensures that the computation cannot be efficiently offloaded or emulated on non-resident memory (e.g., system DRAM or CPU memory).  
In doing so, the VRAM residency primitive functions as a non-intrusive, on-demand audit signal that triggers only under contention, enabling energy-aware, passive compliance detection.

Specifically, we employ \texttt{Argon2id} in keyed mode, where its complexity stems from the properties of the underlying \emph{Pseudorandom Function (PRF)} in this case, the BLAKE2b compression function~\cite{aumasson2013blake2}.  
The PRF-based memory indexing, combined with data-dependent memory mixing, forces each access to depend on the previous block's state, creating a dataflow that is inherently bandwidth-bound and resistant to parallel precomputation.  
By doing so, the scheme ties performance directly to actual VRAM access latency, thereby providing a measurable signal of memory residency and making any attempt at offloading or replay computationally and temporally infeasible.

\begin{figure}[t]
    \centering
    \includegraphics[width=0.85\linewidth]{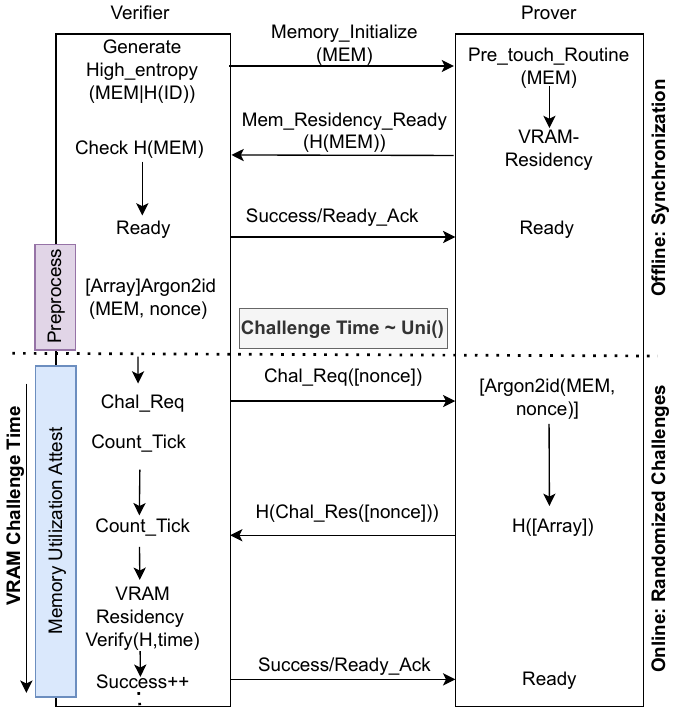}
    \vspace{5pt}
    \caption{Proposed VRAM residency measurement via CRP}
    \label{fig:vram_protocol}
\end{figure}

The core idea is that \textit{hot (resident)} and \textit{cold (non-resident)} memory accesses produce distinguishable timing differences, even observable remotely over a network connection.  
By operating on a large random dataset (\textsc{CHAL}) stored in GPU VRAM (i.e., HBM), the challenger can determine whether the worker's memory resides on-device or has been evicted off-chip, and detect any misuse of HBM for unauthorized use.
As in Figure~\ref{fig:vram_protocol}, first, a pre-challenge is issued to initialize the challenge dataset in GPU VRAM. 
At a random time, a challenge containing fresh nonces is sent to the GPU worker, which computes and returns a keyed \texttt{Argon2id} response.  
The challenger then checks both the correctness of the response and the timing consistency against the expected latency window.  
Successful verification indicates that the challenge data remained locally resident in VRAM throughout the measurement window. 

\subsection{Measurement via Challenge-Response}

We design a measurement strategy to detect whether a large random dataset (\textsc{CHAL}) remains in GPU-local VRAM or incurs latency from access over PCIe or NVLink.
The key insight is that resident memory access is significantly faster, with latency gaps $\Delta T \approx S/B_{\text{pci}}$ emerging under fine-grained, randomized access patterns that defeat caching.
To extract this signal, we use a two-phase challenge: a nonce-keyed BLAKE2b masking pass followed by Argon2id enforcing randomized access.
This design ensures that relocation of \textsc{CHAL} to slower memory produces an observable delay.
We record both wall-clock \texttt{total\_time} and device-only \texttt{kernel\_time}; their difference reflects residency. 

The approach is intentionally bandwidth-bound, prioritizing memory access over arithmetic throughput, and it provides a statistical inference signal that complements the timing-based telemetry in Section~\ref{sec:approach} by extending observability from compute-bound to memory-bound behavior.
Its key architectural properties are unchanged: it relies on fine-grained, random access patterns that negate compression, reordering, or speculative execution; it incorporates per‑challenge randomness through a nonce to allow unpredictability, it induces bandwidth-bound behavior that yields large, measurable $\Delta T$ gaps whenever \textsc{CHAL} is not resident, and it produces consistent latency differences across repeated resident and cold measurement modes.


\noindent\textbf{Challenge Scheduling for Residency Inference. }
To preserve the integrity of the residency measurement, challenges must arrive at random times, ensuring that the worker cannot anticipate when the dataset \textsc{CHAL} will be inspected.
In our setup, the challenger samples a waiting interval $\tau$ from a uniform distribution $\tau \sim \mathrm{Uniform}(0,\,T_{\max})$, 
where $T_{\max}$ defines the maximum spacing between successive checks.
After each interval $\tau$, a fresh challenge is issued, and the worker's response time is recorded. 
The process then immediately resamples a new $\tau$, repeating this randomized schedule throughout the measurement period. 
This approach ensures full coverage over the interval $[0,T_{\max}]$ without introducing long gaps that an adversary could exploit.
Uniform sampling prevents predictability, keeping \textsc{CHAL} pinned in VRAM for the entire measurement period, letting a cold-access penalty reveal eviction or re-purposing.
Although alternatives like exponential or geometric sampling may simulate memoryless renewal processes, we find no practical advantage over uniform timing for this inference task.

\subsection{Fingerprints for GPU-class Binding} 
\label{sec:fingerprint}

To prevent outsourcing of \textsc{CHAL} to unapproved compute platforms, we adopt a
finite-precision, architecture-dependent fingerprinting mechanism that functions as a \emph{GPU-class signature}.  
This idea is supported by the findings of Clifford~\emph{et al.}~\cite{clifford2025locking} and
Schlögl~\emph{et al.}~\cite{schlogl2023causes}, who show that floating-point operators on modern GPUs produce stable, architecture-specific numerical deviations arising from cuBLAS/cuDNN kernel selection, reduction ordering, and fused-multiply-add behavior.  
Although these deviations do not distinguish individual GPUs, they uniquely characterize a GPU
\emph{model}, enabling us to rule out cross-architecture outsourcing while allowing outsourcing to
identical GPUs of the same model, which is acceptable under our threat model.

Given a fixed inference subroutine and a canonical input, the worker computes a
GPU-class fingerprint tied to the numerical behavior of its execution environment.  
The fingerprint is bound to the measurement transcript and tied to the model instance used to solve
\textsc{CHAL}, ensuring that the worker executes both the correct model and the correct hardware
class. 
The preprocessing transforms \textsc{CHAL} using two deterministic phases:
(1) extraction of a GPU-class fingerprint from floating-point operators, and  
(2) a masking transformation over \textsc{CHAL} derived from this fingerprint.  
Both phases must be executed locally on the worker's GPU, forcing full-bandwidth VRAM traversal and
preventing precomputation. 

\noindent\textbf{Floating-Point Drift as GPU-Class Signal.}
To further strengthen memory-residency measurements, we explore the use of a GPU-class fingerprint based on finite-precision floating-point drift.
A fixed sequence of linear layers is evaluated under multiple batched reshapes, each triggering distinct kernel paths in the cuBLAS/cuDNN stack.
The accumulated output deviations yield an architecture-dependent error vector $E$, which remains stable across devices of the same model but diverges across heterogeneous accelerators~\cite{clifford2025locking}.
This vector is then digested into a compact identifier $R_{\mathrm{GPU}} = \mathrm{BLAKE2b}(\mathrm{encode}(E))$.

To tie this signal to memory activity, we use $R_{\mathrm{GPU}}$ as a key in a BLAKE2b-based pseudorandom mask that is applied across all blocks of \textsc{CHAL}, producing $\textsc{CHAL}_{\mathrm{GPU}}[j] = \textsc{CHAL}[j] \oplus \mathrm{mask}_j$.
This keyed masking forces a linear scan of \textsc{CHAL} in VRAM and embeds a device-class dependency into the transformation, which, if recomputed on a mismatched GPU, would induce timing drift or verification error.
While not a secure binding, this measurement offers an additional knob for architectural coupling in residency inference. 

\section{Evaluations}\label{sec:results}
\subsection{Implementation and Experimental Setup}
We implemented the proposed measurement framework using a hybrid Python–CUDA setup, in which the Python controller handles challenge generation, timing instrumentation, and response verification, while all measurement computations are executed on the GPU.
All GPU kernels were compiled with \texttt{nvcc} under CUDA12.4\cite{nvidia_cuda_programming_guide_2025} and tested on NVIDIA Turing and Hopper architectures.

For PoW and VDF workloads, the host controller issues challenges, dispatches GPU kernels, and uses asynchronous CUDA events to measure response latency.
The VDF construction leverages NVIDIA's \texttt{CGBN} (CUDA Big Number) library~\cite{nvidia_cgbn_2025} to support modular exponentiation and large-integer arithmetic within a sequential squaring loop.
Each instance maintains strict sequentiality while exploiting GPU parallelism across multiple VDF chains. 
For GEMM-based puzzles, we used NVIDIA's \textit{CUTLASS}~\cite{cutlass} to implement matrix-multiply–accumulate (MMA) operations targeting tensor cores.
To support the VRAM residency measurement, we developed a custom memory-optimized CUDA implementation of the \texttt{Argon2id} hash function~\cite{biryukov2021argon2} in keyed mode.
This kernel employs BLAKE2b-based compression functions~\cite{aumasson2013blake2} with randomized, data-dependent indexing to induce bandwidth-bound memory access over the incompressible challenge dataset stored in GPU-local memory. 

\subsection{Experimental Parameters}

For reproducibility and clarity, we summarize the main parameters used throughout our evaluation.
For the PoW measurement, we used a difficulty parameter $p$ that targets approximately one valid hash per $2^{24}$ trials, using fresh salts for each solution to ensure independence of inter-solution times.
For GEMM-Based Tensor-Core measurement, we applied Freivalds' algorithm~\cite{freivalds1977probabilistic} with $k = 5$ rounds for verification, using FP16 multiply and FP32 accumulation, the default tensor-core mode.
The VRAM residency experiments leveraged Argon2id flexibility, where we configured 1 pass, 1 lane, and 1\,MiB of working memory per instance. Challenge times were drawn from a uniform distribution over 120\,s for non-continuous probing. 
\textit{Hot} and \textit{cold} residency cases were evaluated by explicitly forcing placement in HBM versus pinned host memory. Network-facing challenge/response traffic (for telemetry rounds) traversed the VM's standard GCP VPC networking stack.



\subsection{Reliability Analysis}

In this section, we evaluate the reliability of the proposed GPU utilization detection mechanisms.  
Specifically, we analyze the response consistency and timing stability of the four measurement schemes, PoW, VDF, GEMM,  and VRAM residency, under varying load and network conditions.

\begin{figure}[t]
\centering
\includegraphics[width=0.8\columnwidth]{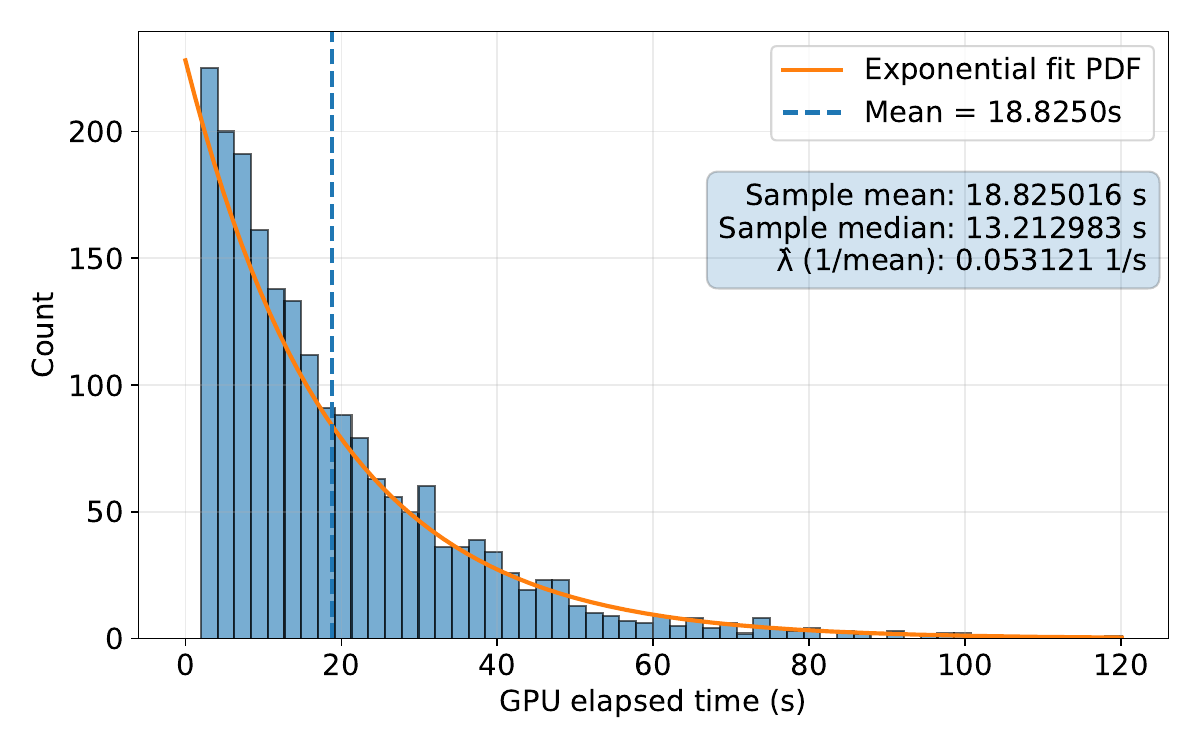}
\caption{Histogram of resolution time of a PoW puzzles for N=1000 solutions and Difficulty = $2^{13}$ }
\label{fig:gpu_single_diff_hist}
\end{figure}

\noindent\textbf{Response Time in PoWs}
Figure~\ref{fig:gpu_single_diff_hist} shows the histogram of solving times for the constructed PoW puzzle executed on an NVIDIA T4 GPU. In our implementation, we set the used block sizes to 32 and set the memory requirement of the \texttt{Argon2id} as 1MB per thread. The calculation is configured for a single lane, and the time complexity is $p=1$ per instance.   
It can be clearly observed that the distribution of solving times follows an exponential trend, consistent with the theoretical model.  

\begin{figure}[t]
\centering
\includegraphics[width=0.8\columnwidth]{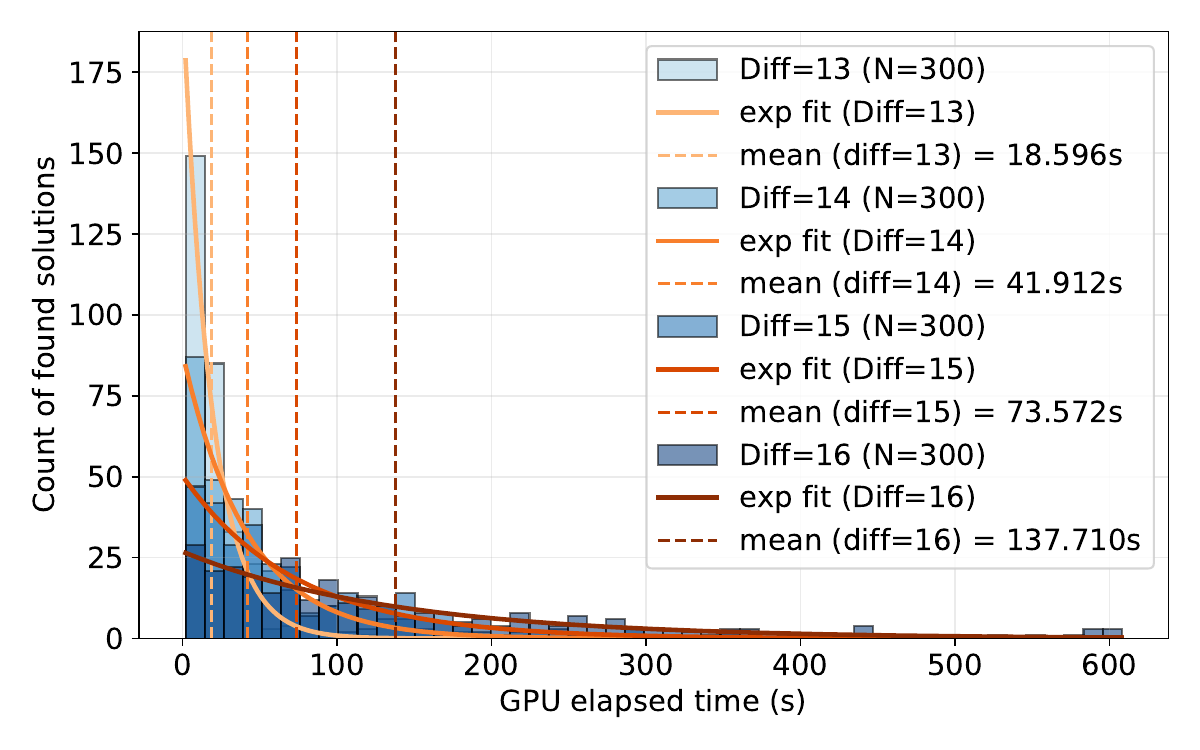}
\caption{Histogram of resolution time of a PoW puzzle for N=300 solutions and different difficulties}
\label{fig:gpu_diff_shift}
\end{figure}

As the puzzle difficulty increases, the corresponding resolution time also grows proportionally, as illustrated in Figure~\ref{fig:gpu_diff_shift}.
Each increment in difficulty results in an exponential increase in the average solution time, reflecting the expected behavior of the PoW process, where success probability decreases exponentially with tighter target thresholds.  
This property enables the measurement scheme to support \textbf{dynamic configurability}, allowing the challenger to adjust the challenge difficulty according to measurement policies and desired detection latency.

\begin{figure}[t]
\centering
\includegraphics[width=0.8\columnwidth]{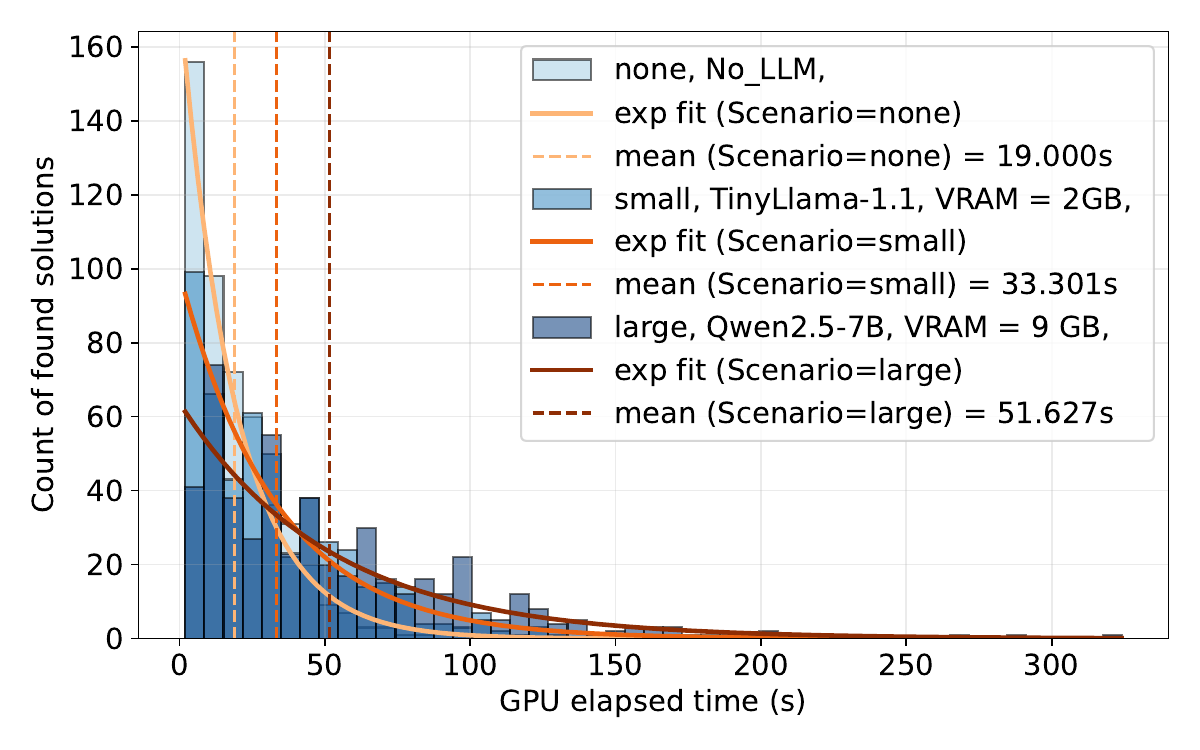}
\caption{Effect of GPU resource contention on PoW response time across different workload scenarios (Difficulty = $2^{13}$ and N=300)}
\label{fig:gpu_llm_contention}
\end{figure}

\begin{figure}[t]
\centering
\includegraphics[width=0.8\columnwidth]{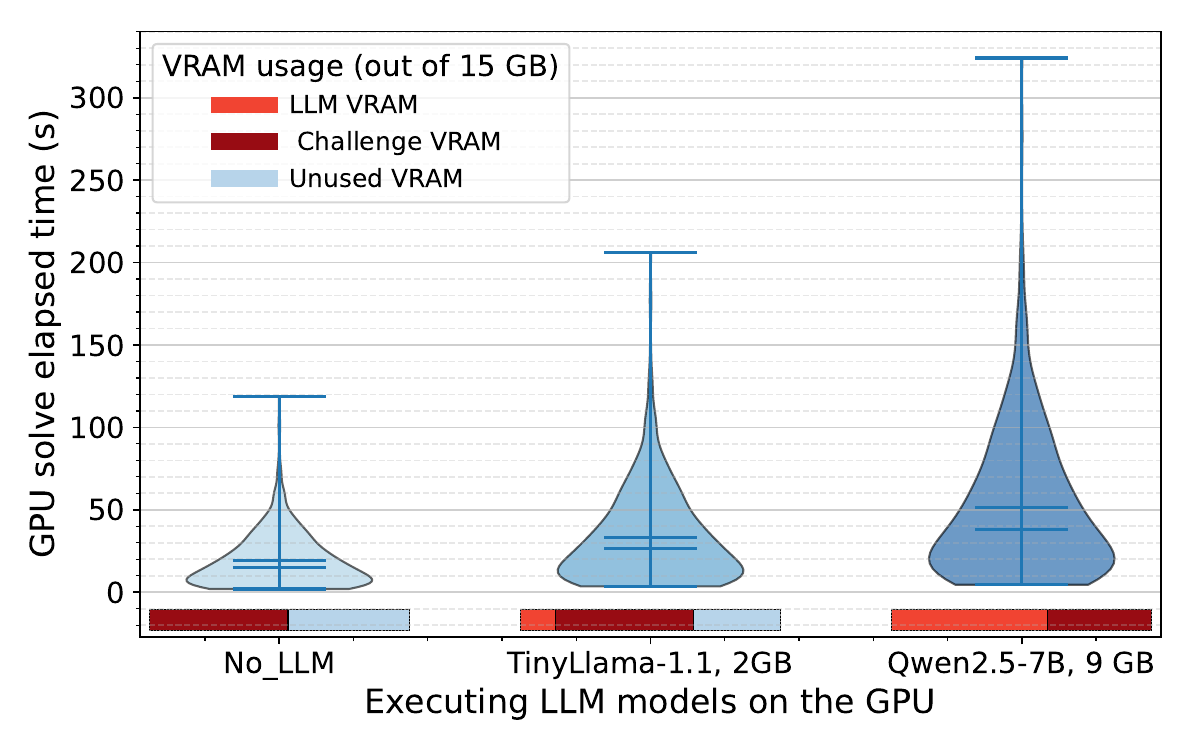}
\caption{Violin distribution of GPU resource contention on PoW response time across different workload scenarios (Difficulty = $2^{13}$ and N=300)}
\label{fig:violin_by_diff}
\end{figure}

In another experiment, we investigate the effect of GPU resource contention caused by concurrent large language model (LLM) workloads on the PoW response timing.  
Specifically, we evaluate three scenarios: (1) no LLM workload, (2) a lightweight \texttt{TinyLlama-1.1} model utilizing approximately 2\,GB of GPU memory~\cite{tinyllama1.1b_ollama}, and (3) a large \texttt{Qwen2.5-7B} model occupying about 9\,GB of T4 GPU memory~\cite{qwen2.5-7b_ollama}.  
In each case, the PoW measurement mechanism operates concurrently with the LLM inference process on the same GPU.  
Figure~\ref{fig:gpu_llm_contention} and Figure~\ref{fig:violin_by_diff} present the observed response timings.  
As predicted, the larger model incurs a higher delay due to increased contention for shared GPU compute and memory resources, whereas the smaller model exhibits a moderate delay relative to the baseline scenario without LLM interference.  
These results demonstrate the sensitivity of the proposed PoW-based measurement to actual GPU utilization, confirming its potential as a fine-grained utilization monitor under varying AI workloads.

\begin{figure}[t]
    \centering

    \begin{subfigure}{0.4\columnwidth}
        \centering
        \includegraphics[width=\linewidth]{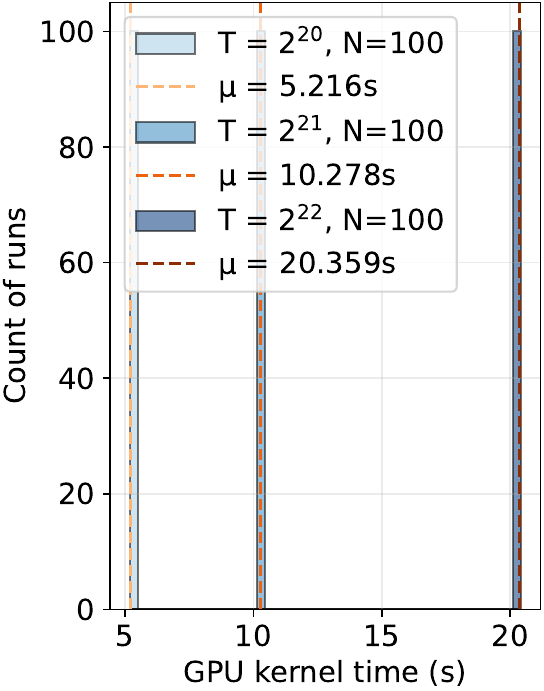}
        \caption{}
        \label{fig:kernel_hist}
    \end{subfigure}
    \begin{subfigure}{0.4\columnwidth}
        \centering
        \includegraphics[width=\linewidth]{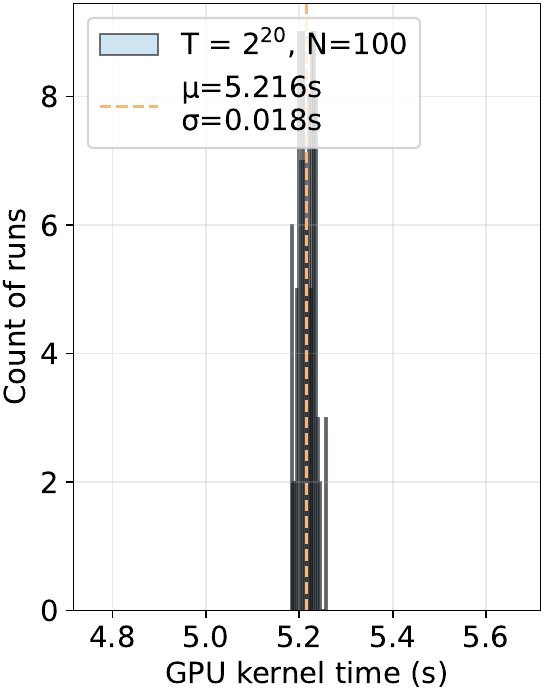}
        \caption{}
        \label{fig:power_hist}
    \end{subfigure}
    \vspace{10pt}
    \caption{T4 GPU kernel response time for M=256 instance of VDF puzzles across different puzzle steps(T).}
    \label{fig:gpu_tlp_dual}
\end{figure}

\noindent\textbf{Multi-Instance VDF Delay Analysis}
For the multi-instance VDF measurement method, we conduct several timing analyses.  
Figure~\ref{fig:gpu_tlp_dual} shows the solve-time distributions for $M=256$ instances, each sampled $N=100$ times across different difficulty parameters $T$.  
As illustrated, VDF-based puzzles exhibit far tighter and more predictable timing behavior than PoW puzzles, which typically suffer from large variance due to their randomized nature. Moreover, the visible step-wise shifts in the distributions reflect the near-linear growth in solve time as the difficulty parameter $T$ increases. 

As another experiment, we vary the number of parallel VDF instances $M$ and record the resulting batch runtime $t_{\mathrm{VDF}}(M)$ to observe saturation effects.  
We define a normalized utilization proxy as $  \mathsf{Util}(M) \;=\; \frac{M / t_{\mathrm{VDF}}(M)}{\max_{M}\big(M / t_{\mathrm{VDF}}(M)\big)},$
which captures the relative throughput at each $M$ with respect to the empirically observed peak.





 We also examine how concurrent LLM workloads affect the solve-time distribution of the multi-instance VDF measurement. Unlike PoW puzzles, which stress memory bandwidth, VDF evaluation is dominated by big-integer modular squaring executed on CUDA cores. Consequently, any workload that competes for these same scalar units directly influences VDF latency. 

\begin{figure}[t]
\centering
\includegraphics[width=0.8\columnwidth]{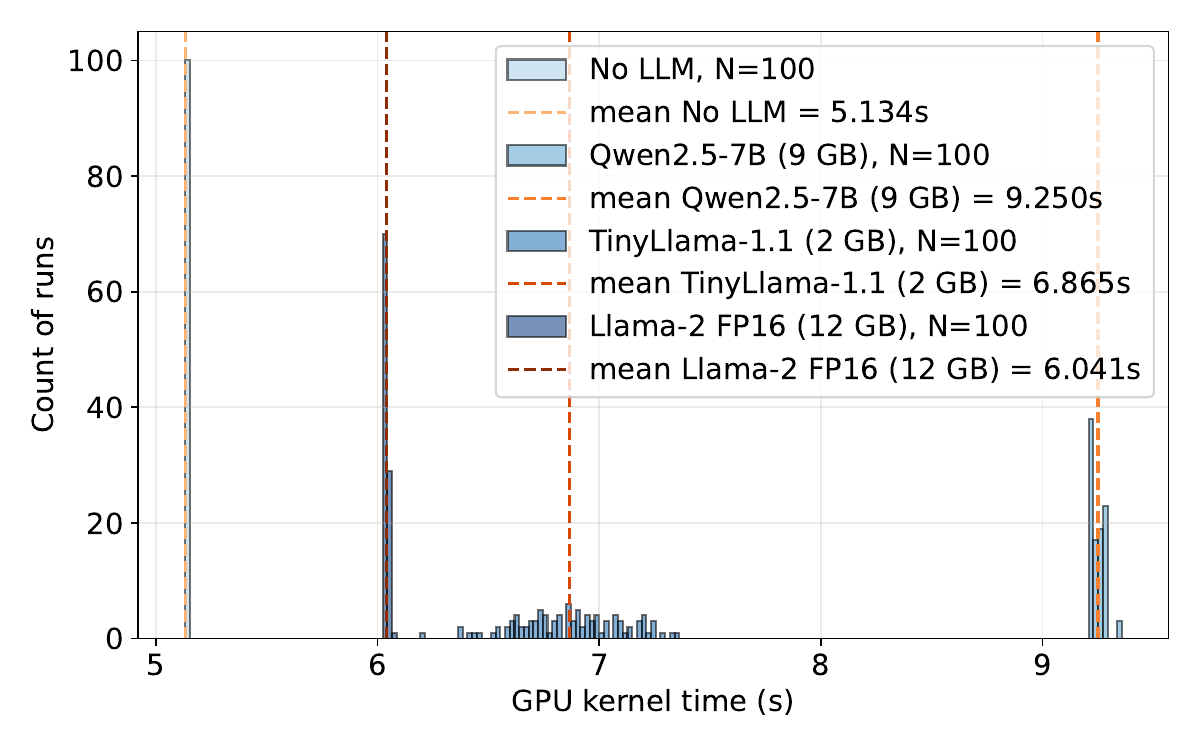}
\caption{Response time histogram of M=128 VDF instances for $T=2^{20}$ in contention with LLM models.}
\label{fig:tlp_kernel_hist_scenarios}
\end{figure}

We evaluate the same three scenarios in the PoW experiment with the addition of a \texttt{Llama-2 FP16} model occupying approximately 12\,GB of GPU memory~\cite{touvron2023llama}. The models are executed in contention with M=256 instance of VDF on $T=2^{20}$ steps. Importantly, the models differ in their dominant compute pathways, where some kernels primarily utilize tensor cores while others rely on CUDA cores (depending on the implementation), resulting in distinct contention patterns on the T4 GPU. Figure~\ref{fig:tlp_kernel_hist_scenarios} shows the resulting VDF timing histogram. These results highlight that VDF-based measurement is highly sensitive to contention on the CUDA core pipeline, enabling it to capture real-time utilization fluctuations even under mixed AI workload conditions.

\begin{figure}[t]
\centering
\includegraphics[width=0.8\columnwidth]{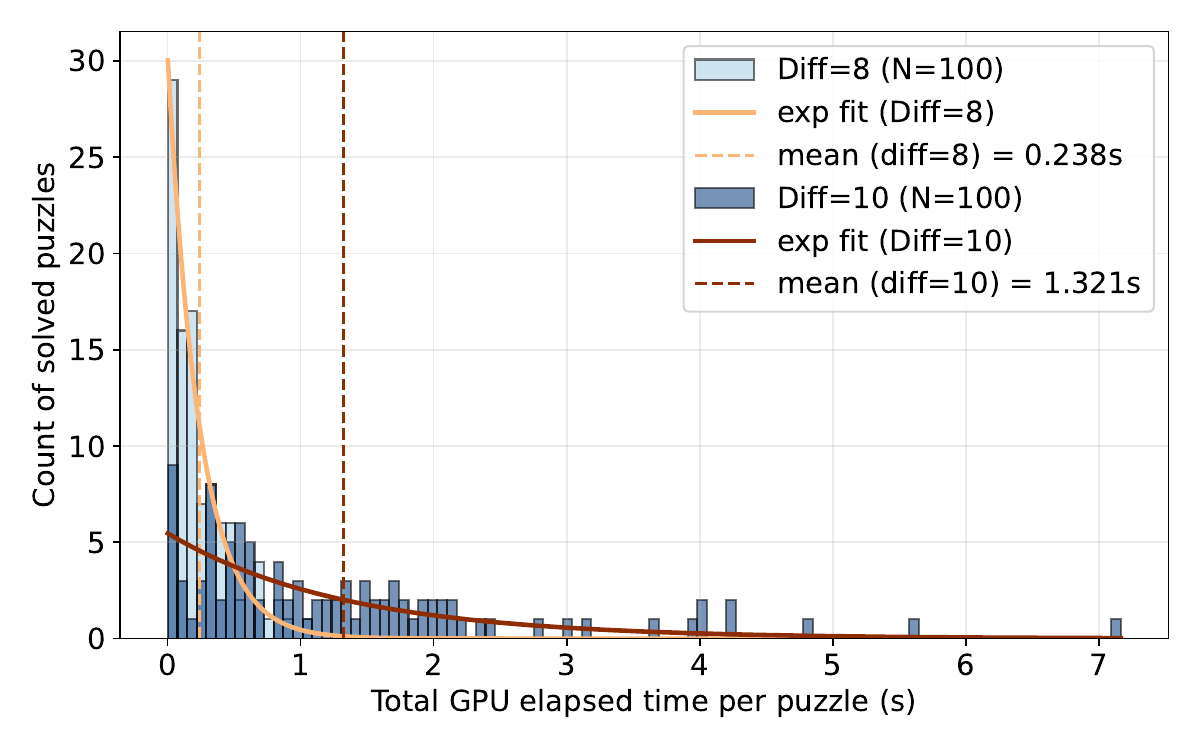}
\caption{Solutions response time histogram distribution on H100 based on GEMM-equipped PoW puzzles across 2 difficulties. The dimension of the square matrices in GEMM is fixed at $N=2^{15}=32,768$}
\label{fig:gemm_gpu_times_by_difficulty}
\end{figure}

\noindent\textbf{Response Time in GEMM Puzzles. }\label{sec:gemm_res}
The next set of experiments is conducted with regard to the GEMM-based PoW framework, which is designed to involve the tensor-cores on the target GPU. For these set of experiments we have taken into account the H100 GPU as our primary target. Figure~\ref{fig:gemm_gpu_times_by_difficulty} depicts the exponential distribution of resolution time by the GPU for the probabilistic puzzles similar to the PoW approach, but with the underlying GEMM-based operations instead of hash calculation.  

\begin{figure}[t]
\centering
\includegraphics[width=0.77\columnwidth]{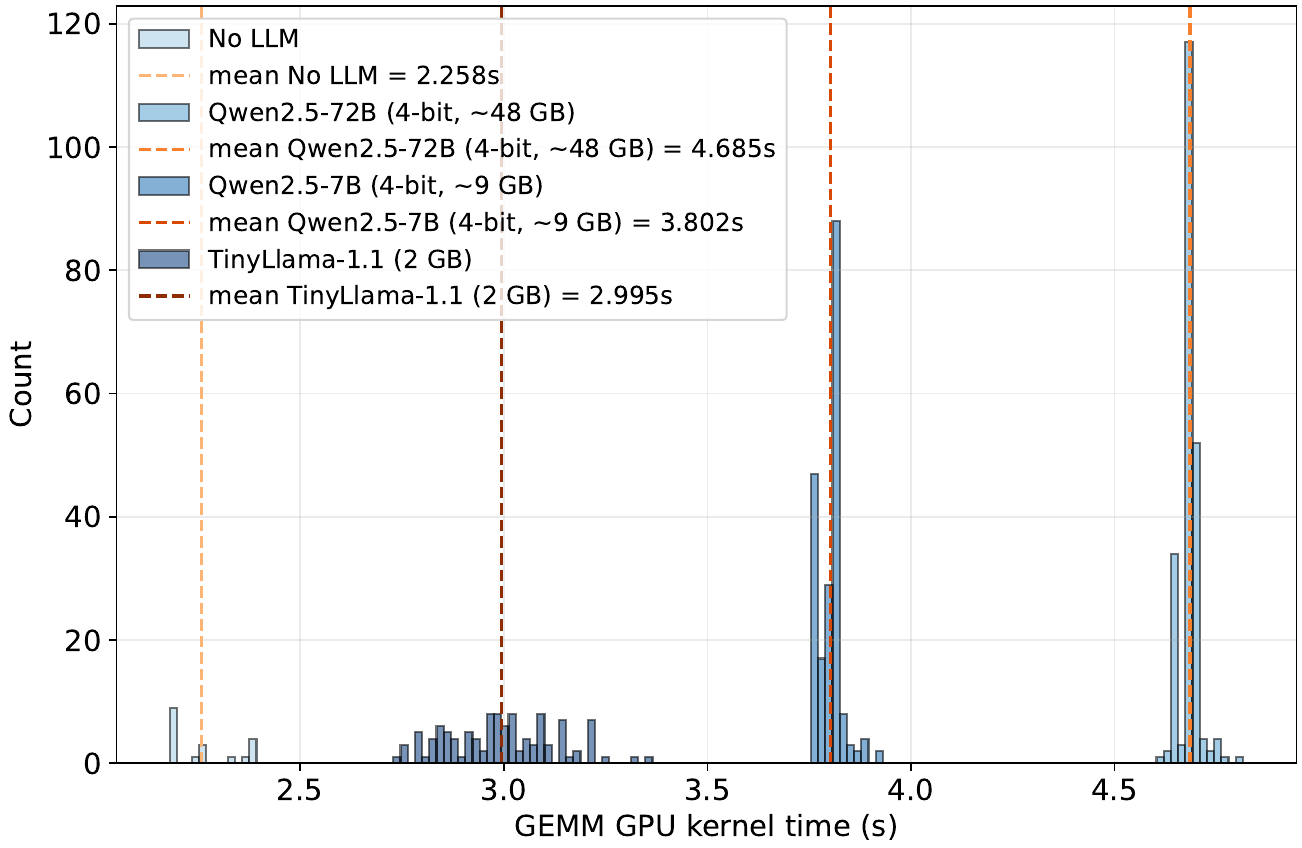}
\caption{Histogram of GEMM-PoW response times for $Diff=10$ and $N=2^{15}$ under LLM-induced contention.}
\label{fig:llm_contention_gemm}
\end{figure}

Finally, we evaluate the same LLM-contention experiment using our GEMM-based PoW construction.
Figure~\ref{fig:llm_contention_gemm} presents the solve-time distributions on an H100 GPU when the prover executes matrix-multiplication puzzles concurrently with different LLM workloads. In contrast to VDFs, which are sensitive to pressure on CUDA  units, the GEMM-PoW is entirely dominated by tensor-core execution. As a result, models with heavier tensor-core utilization impose substantially greater slowdown on PoW evaluation.

As the Figure illustrates, larger models introduce proportionally larger delays, reflecting intensified contention in the tensor-core pipeline and shared-memory/throughput subsystems used by both GEMM and FP16/BF16 LLM inference. 
These results demonstrate that GEMM-PoW measurement captures real-time contention effects originating specifically from tensor-core–bound workloads, allowing us to isolate and measure activity in the deep-learning execution path independently from the CUDA-core path.

\begin{figure}[t]
\centering
\includegraphics[width=0.8\columnwidth]{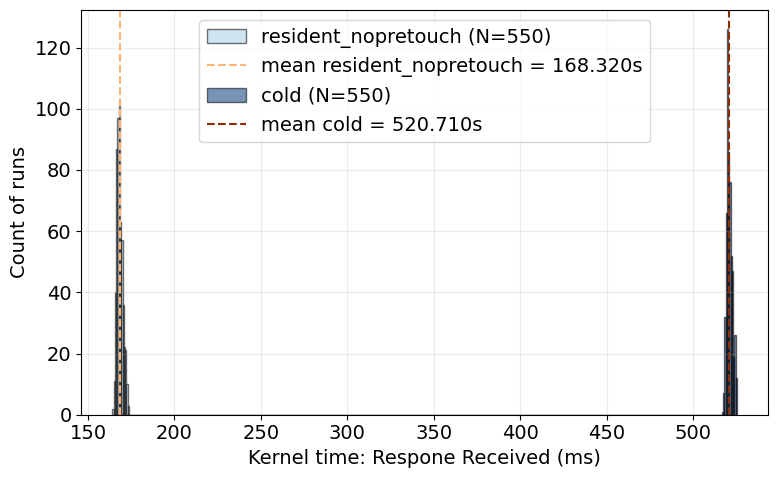}
\caption{Response time histogram of N = 550 test VRAM Residency measurements on H100 for \textit{CHAL} = 60 GB.}

\label{fig:residency_local}
\end{figure}

\noindent\textbf{VRAM Residency Delay Analysis. }
Before evaluating the delay-based residency test, we clarify the memory allocation behavior on modern GPUs cf.~\cite{nvidia_cuda_programming}. 
When a large challenge buffer (CHAL) is allocated, its residency depends on whether it is backed by GPU-local HBM or resides in host-pinned system RAM accessed over PCIe. 
In our tests, ``hot'' challenges refer to buffers explicitly allocated in HBM (e.g., via cudaMalloc or by first clearing the PyTorch caching allocator via torch.cuda.empty\_cache() to force a fresh device allocation). 
Conversely, ``cold'' challenges reside in host-pinned memory but are mapped for direct device access (e.g., via cudaHostAlloc or PyTorch’s pin\_memory()). 
This physical residency governs access latency, with HBM providing significantly higher bandwidth than the PCIe interconnect, forming the basis of our residency detection mechanism.
By analyzing the statistical distribution of response delays, we show that our mechanism provides stable and consistent detection, demonstrating that the measurement framework remains robust in HBM-equipped GPU architectures and realistic execution environments.
Figure~\ref{fig:residency_local} presents the histogram for the VRAM residency test under two conditions: a \textit{hot} scenario where the \textit{CHAL} resides entirely within the GPU's HBM, and a \textit{cold} scenario where \textit{CHAL} must be fetched from pinned system RAM. As clearly illustrated, the timing gap between the two modes exceeds 350 ms, making them trivial to distinguish.

\begin{figure}[t]
\centering
\includegraphics[width=0.83\columnwidth]{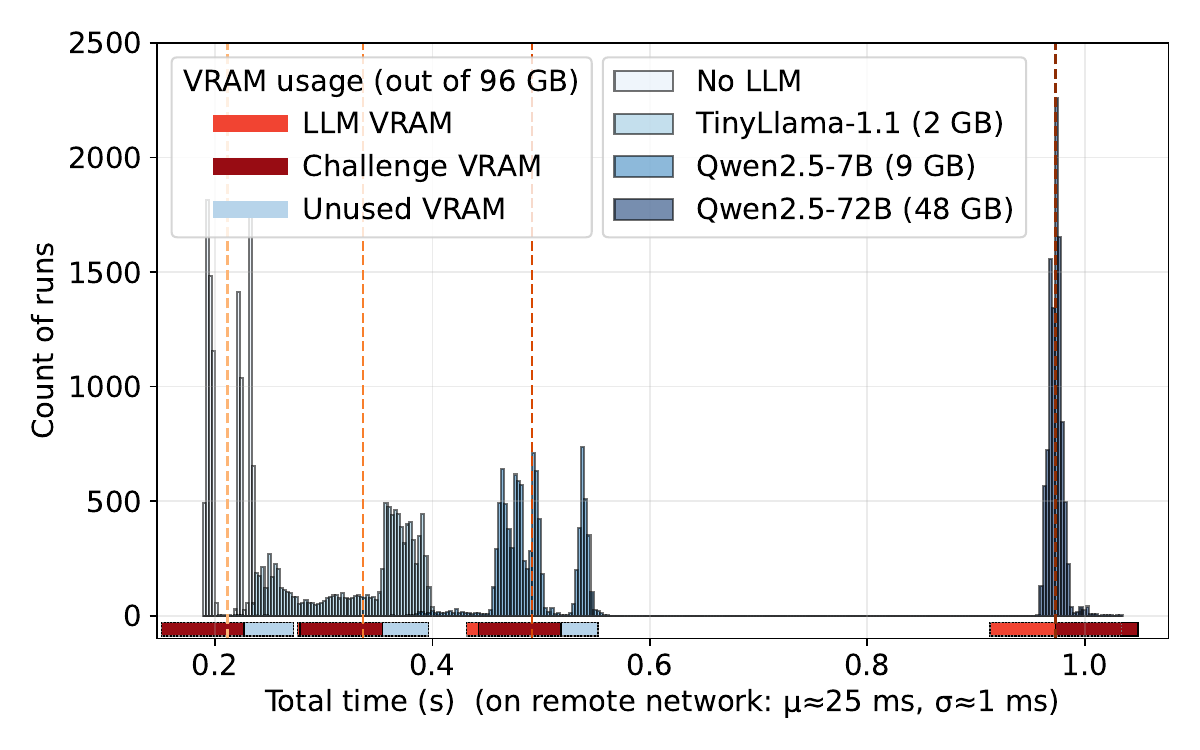}
\caption{H100 Response time histogram of VRAM proximity method with VRAM \textit{CHAL} = 60 GB in contention with LLM models.}
\label{fig:residency_llm}
\end{figure}

We further evaluate the \textit{VRAM Residency} test under contention with multiple LLM workloads of varying memory footprints. Since LLM training and inference parameters reside primarily in HBM, the residency of the predetermined \textit{CHAL} directly competes with model parameters depending on model size. Figure~\ref{fig:residency_llm} shows the resulting response-time distributions on an H100 GPU. 

For the largest model (Qwen2.5-72B), the memory footprint exceeds the GPU's available HBM capacity, forcing part of the \textit{CHAL} to spill into host RAM. In this case, the challenge response becomes \textit{cold}, incurring a visibly larger delay compared with smaller models, as shown in the Figure. Consequently, excessive HBM pressure becomes immediately detectable whenever the device is challenged.
To successfully pass the residency test, a user must keep the \textit{CHAL} hot in HBM, an action that significantly reduces the model's effective throughput (tokens processed per second). This creates a built-in deterrent, ensuring that any attempt to hide intensive model execution results in measurable performance degradation. 

\begin{figure}[t]
\centering
\includegraphics[width=0.8\columnwidth]{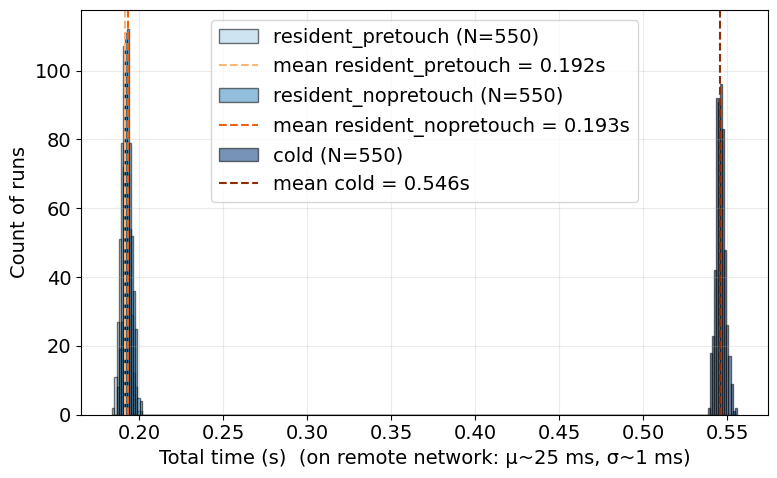}
\caption{Response time histogram of VRAM residency measurements under remote execution and pre-touch conditions.}
\label{fig:residency_internet}
\end{figure}

We also evaluate VRAM residency detection under remote-execution conditions and with pre-touch optimizations. Figure~\ref{fig:residency_internet} compares timing distributions when the challenge buffer is pre-touched versus accessed normally from HBM. Despite reduced latency in the pre-touch case, the distinction between HBM-resident and non-resident challenges remains clearly detectable.

\begin{figure*}[t]
    \centering
    \begin{subfigure}[t]{0.323\textwidth}
        \centering
        \includegraphics[width=\linewidth]{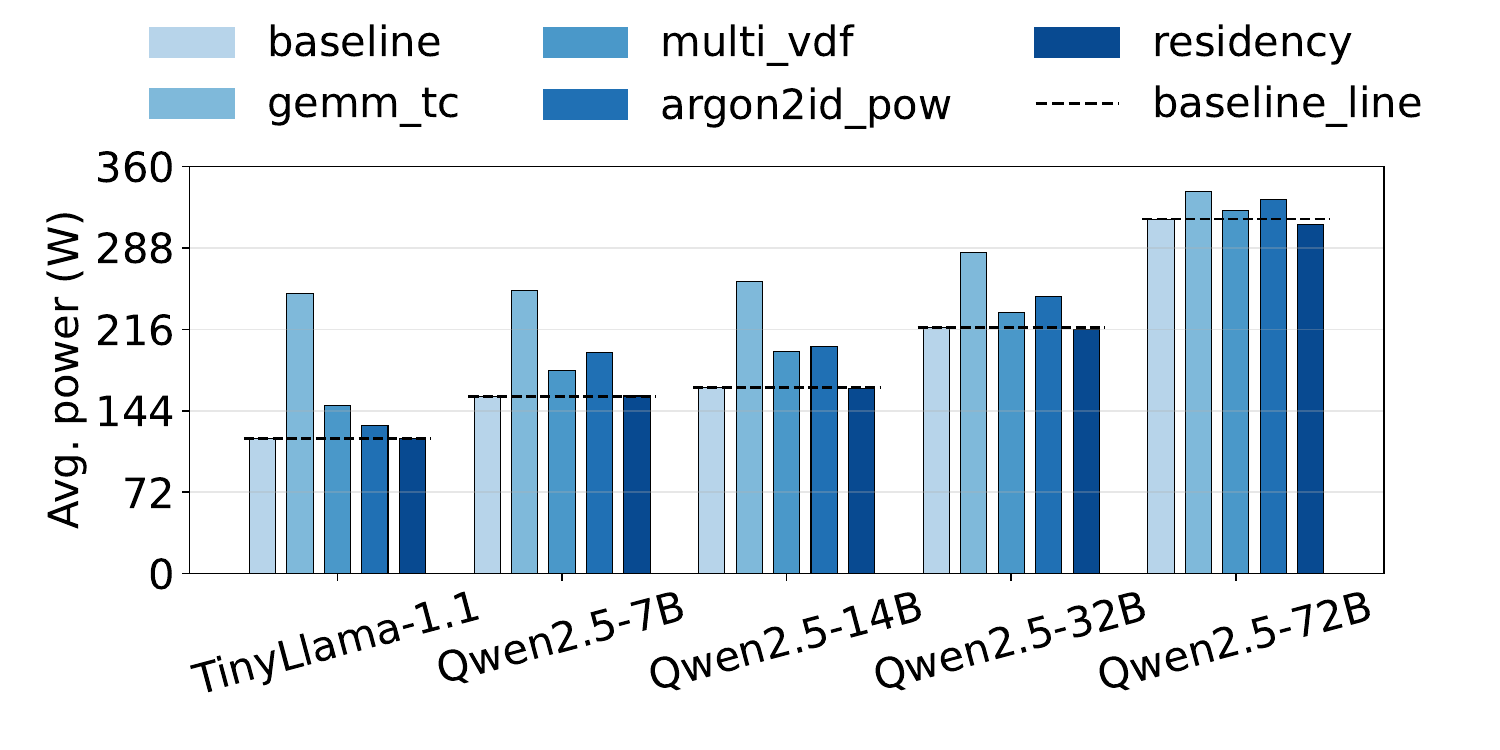}
        \caption{Average Power Consumption}
        \label{fig:plotC}
    \end{subfigure}\hfill
        \begin{subfigure}[t]{0.323\textwidth}
        \centering
        \includegraphics[width=\linewidth]{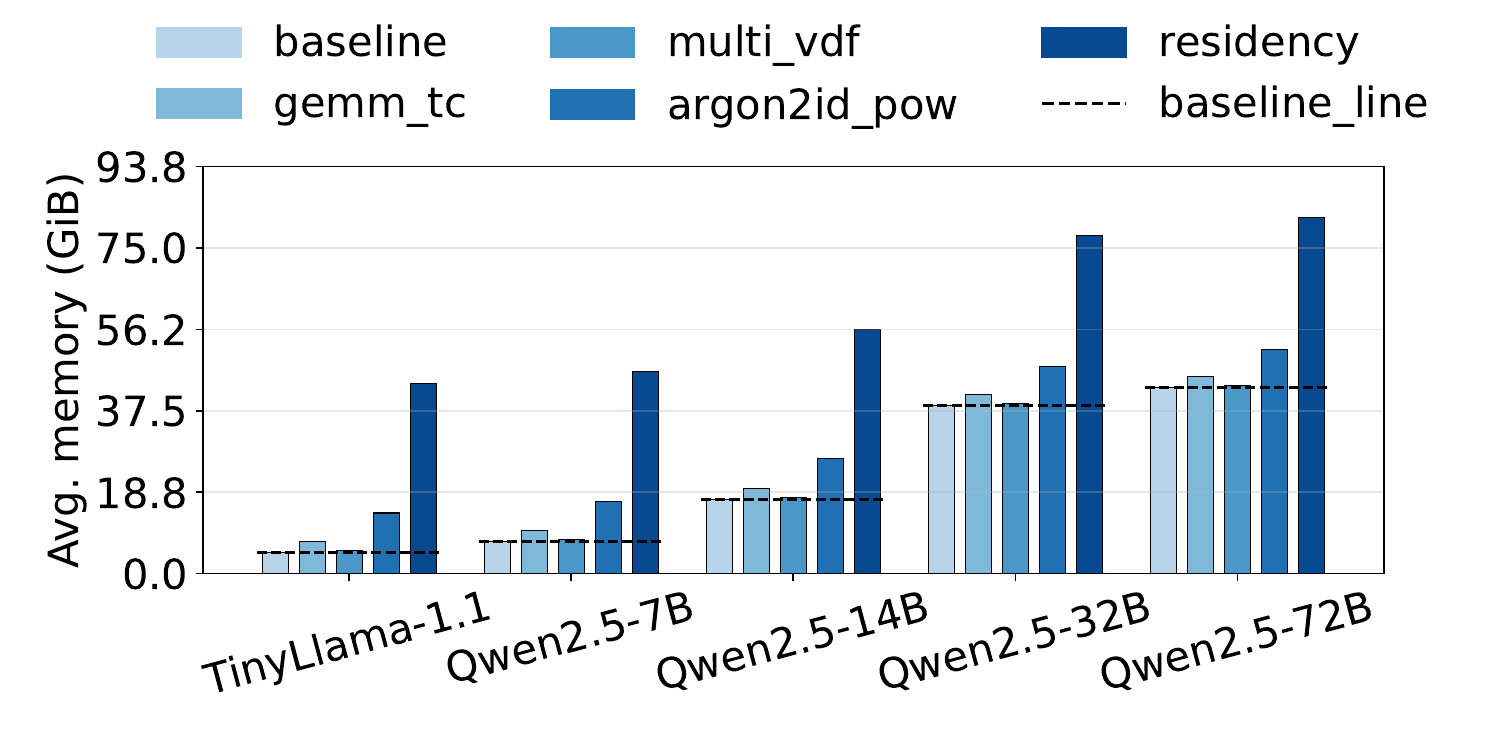}
        \caption{Average Memory Consumption}
        \label{fig:plotB}
    \end{subfigure}\hfill
    \begin{subfigure}[t]{0.323\textwidth}
        \centering
        \includegraphics[width=\linewidth]{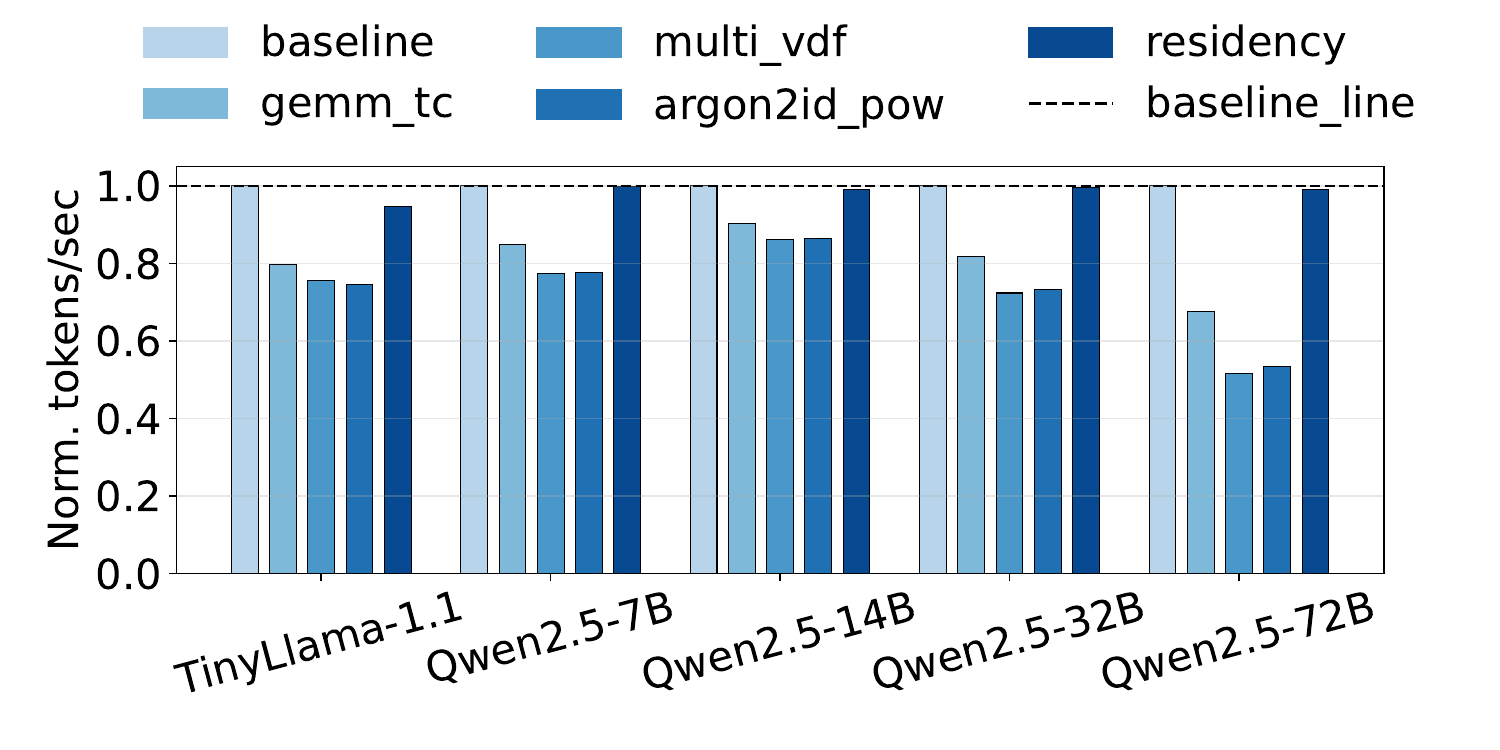}
        \caption{Normalized Token/Sec Throughput}
        \label{fig:plotA}
    \end{subfigure}\hfill

    \vspace{4mm}
        \caption{Overhead comparison between baseline execution and four measurement mechanisms over 10-min interval.}

    \label{fig:three_wide_subfigs}
\end{figure*}
\subsection{Overhead Analysis}

To quantify the operational cost of our measurement mechanisms, we evaluate four different approaches \textit{GEMM-TC}, \textit{Multi-instance VDF}, \textit{Argon2id-PoW}, and \textit{VRAM Residency} together with a \textit{baseline} scenario in which no measurement is performed.  
All results were collected over a 10-minute evaluation window, during which the LLM workloads (described in Section~\ref{sec:gemm_res}) remained continuously active.  
For the \textit{VRAM Residency} method, measurement events were triggered at randomized intervals following a uniform distribution over two minutes, 
reflecting the intended design where memory-residency checks occur unpredictably rather than continuously.
Figure~\ref{fig:three_wide_subfigs} summarizes the resulting overheads in terms of (a) average power draw, (b) average memory usage, and (c)  normalized token throughput.

\noindent\textbf{Power Overhead. }
Figure~\ref{fig:plotC} highlights that continuous- measurement \textit{GEMM-TC}, \textit{Multi-instance VDF}, and \textit{Argon2id-PoW}, introduce non-trivial power overhead compared to the baseline.  
This is expected as each method requires persistent compute activity on CUDA or tensor cores, which elevates GPU power draw across the entire measurement window. 
In contrast, the \textit{VRAM Residency} test exhibits only negligible power overhead.  
This follows from its low-duty-cycle design, in which the residency probe executes only when a randomized challenge is issued, and each probe performs a single hash over a fixed region of VRAM. Between challenges, the GPU remains dedicated to legitimate AI workloads and consumes no additional measurement-related power.

\noindent\textbf{Memory Overhead. }
As shown in Figure~\ref{fig:plotB}, all continuous measurement methods incur significant GPU memory overhead.  
Both \textit{GEMM-TC} and \textit{Multi-instance VDF} allocate working buffers, while \textit{Argon2id-PoW} requires a larger stateful memory for repeated hashing rounds.  
This memory pressure reduces the headroom available for LLM inference and may force memory paging or model quantization in resource-constrained environments. Unlike the other three approaches, the \textit{VRAM Residency} mechanism intentionally incurs substantial memory overhead.  

This follows directly from the proposed design where the \textit{CHAL} region must occupy a large, contiguous portion of the GPU's HBM  so that residency violations can be detected with high sensitivity.  
Consequently, the memory footprint of the \textit{VRAM Residency} test is large by design, even though it performs no additional allocations beyond the \textit{CHAL} itself.

\noindent\textbf{Performance Overhead (Token/Sec). }
Figure~\ref{fig:plotA} reports the normalized token/sec throughput under each measurement strategy.  
For these measurements, we assume a legitimate model is in continuous use, and therefore the performance metric reflects the user-visible impact on actual inference workloads. Continuous-measurement methods (\textit{PoW-Based} and \textit{VDF}) cause noticeable degradation in throughput, primarily because they compete directly with LLM kernels for GPU compute, memory bandwidth, and scheduling slots.
In contrast, \textit{VRAM Residency} incurs negligible throughput loss and does
not disrupt tensor-core execution or GPU scheduling.

\section{Conclusion and Discussion}\label{sec:conclusion}

We introduced a complementary approach to hardware-rooted trust: \emph{compute-based verifiable GPU utilization measurement}.  
Rather than depending on vendor-enforced telemetry, secure enclaves, or proprietary firmware, our framework uses \emph{compute-based evidence} derived from time-dependent cryptographic puzzles and HVM behavior to reveal whether a GPU is actively executing large-scale workloads.

We presented four measurement primitives that jointly cover the major execution pathways of modern accelerators: a probabilistic Proof-of-Work signal, a sequential Verifiable Delay Function for scalar pipelines, GEMM-based tensor-core puzzles, and a bandwidth-sensitive VRAM residency measure that exposes memory locality through measurable latency shifts.  
Our measurements rely on GPU-specific properties; for instance, co-located tensor-core activity or concurrent kernel execution perturbs GEMM, PoW, and VDF solve times in a manner that is statistically detectable, even under optimized schedules.
Similarly, VRAM-based challenges induce a distinct cold-access penalty when dataset residency is violated, enabling lightweight memory usage inference without relying on trusted hardware or telemetry.

While our fingerprinting-based masking mechanism was evaluated in a non-continuous setting, its integration with randomized probes offers a promising direction for future continuous telemetry mechanisms.
Periodic sampling with ephemeral nonces and small Argon2 probes could enforce ongoing residency checks with low overhead, subject to constraints imposed by co-residency, scheduling, and bandwidth fairness.
Designing such extensions without disrupting legitimate workloads remains an open challenge. 
While Section~\ref{sec:results} shows separable timing distributions, defining a hardware-specific formal thresholds left for the future work. 
As such, FP/FN rates are not quantified, though they can be estimated empirically under deployment-specific assumptions. 
A detection-theoretic framework remains important future work. 

We emphasize that our measurements are not designed to deliver cryptographic proof of correct execution. 
Instead, they offer inference primitives that can inform future enforcement strategies based on statistical, behavioral, or temporal patterns.
Integration into future governance protocols remains an open design question and would need to account for adversarial adaptation, privacy leakage, and deployment scalability.  

\noindent\textbf{Protocol-level applications.}
The telemetry primitives introduced in this work—timing distributions, memory residency inference, and class-specific throughput profiles—may serve as foundational signals in future protocol designs. 
While we do not formalize or evaluate such protocols here, we note their potential utility in a range of post-deployment attestation and accountability settings. 
These include:
(1) location estimation, where delay-amplified interactions can help detect geographic misrepresentation;
(2) tenancy discrimination, where telemetry aids in identifying co-resident workloads or covert GPU sharing;
(3) compute fairness and scheduling audits, where runtime observables can detect starvation, overuse, or deviation from promised service levels; and
(4) resource binding, where verifiable execution patterns constrain a workload to a specific accelerator class (e.g., FP8-capable H100 vs. older Volta).
Such protocols would complement existing cryptographic attestation and isolation mechanisms, and may help enforce transparency, policy compliance, or regional restrictions in multi-tenant AI deployments.

\bibliographystyle{ACM-Reference-Format}
\bibliography{references}
\small

\section*{A. PoW Timing Model}
\label{app:pow_model_full}
A PoW puzzle requires the prover to find an input $x$ such that the output of a cryptographic hash function $H(x)$ satisfies a target condition
\begin{equation}
H(x) < \text{target}
\end{equation}
where the target defines the puzzle’s difficulty.
The verifier can efficiently confirm the solution by recomputing $H(x)$, while the prover must perform a brute-force search over many candidate inputs $\{x_i\}$ until the inequality holds, making the expected computational effort proportional to the inverse success probability \(1/p = 1/\Pr[H(x) < \text{target}]\).
We model PoW solve times under both single-thread and multi-thread (GPU) settings, showing the exponential-time behavior and memoryless property that enables a probabilistic detection mechanism for compute utilization.

\noindent\textbf{PoW Puzzle Structure.}
Each hashing attempt is modeled as an independent Bernoulli trial with success probability $p = \Pr[H(\text{salt} \parallel x) < \text{target}]$.
If a thread hashes at rate $r$ (hashes/s), the number of attempts $K$ until the first success follows a geometric distribution. For the vanishingly small $p$ typical in PoW (e.g., $p \approx 2^{-32}$), the solve time $T = K / r$ is well-approximated by an exponential distribution:
\begin{equation}
T \sim \mathrm{Exp}(\lambda), \quad \lambda = r p.
\end{equation}

\noindent\textbf{Parallelism and Summation of Rates.}
With $M$ independent processing units (e.g., GPU threads), the system solve time $T_{\mathrm{GPU}}$ is the minimum of $M$ independent exponential variables. Consequently, the aggregate rate is the sum of individual rates:
\begin{equation}
T_{\mathrm{GPU}} \sim \mathrm{Exp}\left(\sum_{i=1}^M r_i p\right) = \mathrm{Exp}(R p), \quad R = M r.
\end{equation}
The resulting distribution is memoryless, meaning the expected remaining time to find a solution does not depend on how much time has already elapsed:
\begin{equation}
\mathbb{E}\left[ T_{\mathrm{GPU}} - s \mid T_{\mathrm{GPU}} > s \right] = \frac{1}{R p}.
\end{equation}

\noindent\textbf{Salt Freshness and i.i.d. Enforcement.}
To ensure that solve times $T_i$ remain independent and identically distributed (i.i.d.), the challenger must issue a \emph{fresh random salt} immediately upon receiving a valid solution. 
This resets the Poisson process and prevents ``progress carrying'' or pre-computation. 
Without fresh salts, an adversary could parallelize across time (pre-computing future solutions), which would violate the Gamma distribution assumption used in the timing tests.

\noindent\textbf{Detection via Solution Timing.} 
By issuing \emph{fresh random salts} for each challenge, the challenger ensures that inter-solution times $T_i$ remain i.i.d. and exponentially distributed. Let $\lambda = R p$ be the expected solve rate. Then:
\begin{equation} 
T_i \sim \mathrm{Exp}(\lambda), \quad \mathbb{E}[T_i] = \frac{1}{\lambda}. 
\end{equation} 
Summing $n$ such intervals yields the total solve time $S_n$:
\begin{equation} 
S_n = \sum_{i=1}^n T_i \sim \mathrm{Gamma}(n, \lambda). 
\end{equation} 
This enables two forms of inference:

\noindent\textbf{(a) Fixed-Sample Test.} 
We test $H_0: \lambda \ge \lambda_{\min}$. Using the property $2\lambda S_n \sim \chi^2_{2n}$, the acceptance condition is: 
\begin{equation} 
S_n \le \frac{\chi^2_{2n,1-\alpha}}{2 \lambda_{\min}} + n t_0, 
\end{equation} 
where $n t_0$ accounts for cumulative network and processing overhead. Note that simply subtracting $t_0$ from individual samples ($T'_i = \max(T_i - t_0, 0)$) creates a shifted exponential distribution; to remain statistically rigorous, the threshold is adjusted by $n t_0$ to maintain the validity of the $\chi^2$ test. 

\noindent\textbf{(b) Fixed-Time Test.} 
In a fixed window $t$, the number of solutions $K$ follows a Poisson distribution $K \sim \mathrm{Poisson}(\lambda t)$. We accept $H_0$ if $K \ge k_{\text{crit}}$, where $k_{\text{crit}}$ is the lower quantile of the Poisson distribution: 
\begin{equation} 
k_{\text{crit}} = F^{-1}_{\mathrm{Poisson}(\lambda_{\min}t)}(\alpha).   
\end{equation} 

\noindent Here, $(1 - \alpha)$ lower confidence bound on the actual rate $\lambda$ is: 
\begin{equation} 
\lambda_L = \frac{\chi^2_{2K, \alpha}}{2t}. 
\end{equation}

\begin{algorithm}[t] 
\small
\caption{Fixed-Sample PoW Verification (level $\alpha$)} 
\label{alg:fixed-sample} 
\DontPrintSemicolon 
\KwIn{$\lambda_{\min}$, $\alpha$, $n$, latency $t_0$} 
\KwOut{Decision on $H_0: \lambda \ge \lambda_{\min}$} 
$S \gets 0$\; 
\For{$i \gets 1$ \KwTo $n$}{ 
    Issue salt; observe solve time $T_i$\; 
    $S \gets S + T_i$\; 
} 
\tcp{Threshold includes cumulative latency shift} 
$\tau \gets \frac{\chi^2_{2n, 1-\alpha}}{2\lambda_{\min}} + n t_0$\; 
\BlankLine 
\leIf{$S \le \tau$}{ 
    \textbf{ACCEPT} $H_0$ 
}{ 
    \textbf{REJECT} $H_0$ 
} 
\end{algorithm}

\begin{algorithm}[t] 
\small
\caption{Fixed-Time PoW Verification (level $\alpha$)} 
\label{alg:fixed-time} 
\DontPrintSemicolon 
\KwIn{$\lambda_{\min}$, $\alpha$, duration $t$} 
\KwOut{Decision on $H_0: \lambda \ge \lambda_{\min}$} 
\BlankLine 
Run PoW for $t$; count solutions $K$\; 
$\mu \gets \lambda_{\min} t$\; 
$k_{\text{crit}} \gets F^{-1}_{\mathrm{Poisson}(\mu)}(\alpha)$\; 
\BlankLine 
\leIf{$K \ge k_{\text{crit}}$}{ 
    \textbf{ACCEPT} $H_0$ 
}{ 
    \textbf{REJECT} $H_0$ 
} 
\end{algorithm}

\noindent\textbf{Guidelines.}
\begin{itemize}
\item Choose $\lambda_{\min}$ to reflect the slowest tolerable on-GPU hash rate.
\item Significance level $\alpha \in [0.01, 0.10]$.
\item Use fixed-sample test with $n \gtrsim 20$ or fixed-time with $\lambda_{\min} t \gtrsim 10$.
\end{itemize}

We stress that these tools support probabilistic telemetry inference of compute throughput under constrained assumptions, and may assist future enforcement policies when embedded into broader protocols.

\section*{B. VDF Construction Details}
\label{app:vdf_full}
We base our VDF construction on Wesolowski’s scheme~\cite{wesolowski2020efficient}, which builds upon the classical Rivest–Shamir–Wagner (RSW) time-lock puzzle~\cite{rivest1996time}. Although several alternative VDF constructions exist (e.g., Pietrzak’s proof system~\cite{pietrzak2019simple}), the Wesolowski–RSW design offers an ideal balance between simplicity, security, and verification efficiency and provides a minimal proof size as well as a logarithmic verification cost. It has also been adapted by Chia~\cite{cohen2019chia} and Ethereum~\cite{bunz2017proofs} projects.
We describe the group setup, sequential squaring evaluation, Wesolowski proof generation/verification, and the multi-instance GPU scaling with batch verification in the following.

Let $N = pq$ be a strong RSA modulus, where $p = 2p’ + 1$ and $q = 2q’ + 1$ are distinct safe primes. 
We define $G = \mathrm{QR}_N$ as the cyclic subgroup of quadratic residues in $\mathbb{Z}_N^{\ast}$, given by $\mathrm{QR}_N = {, x^{2} \bmod N \mid x \in \mathbb{Z}_N^{\ast} ,}$. 
Because the factorization of $N$ is hidden, $G$ is a group of unknown order, with $\lvert G \rvert = p’ q’$, and computing this order is as hard as factoring $N$. 
This structure is fundamental for enforcing sequential computation, as it prevents an adversary from applying Euler's theorem to shortcut the $T$ sequential squarings via a single exponentiation modulo $\lvert G \rvert$. 
Given a public seed $\mathsf{sid}$, we deterministically derive the generator and delay parameter as $g \leftarrow \mathsf{HashToQR}(\mathsf{sid})$ and $T \leftarrow \mathsf{DeriveDelay}(\mathsf{sid})$. Here, $\mathsf{HashToQR}$ maps the hash of $\mathsf{sid}$ into the quadratic-residue subgroup by computing $(H(\mathsf{sid}) \bmod N)^2 \bmod N$, ensuring that $g \in \mathbb{G}$, while $\mathsf{DeriveDelay}$ deterministically selects the delay parameter $T$ (i.e., the number of sequential squarings) from the hash value to control puzzle hardness. The worker evaluates the VDF by performing $T$ dependent squarings, computing $y = g^{2^T} \bmod N,$
where each iteration follows $x_{i+1} = x_i^2 \bmod N$ for $ 0\leq i\leq T-1$, thereby enforcing strict sequentiality that cannot be parallelized even with massive hardware.

To make the computation verifiable, verier computes a  proof $\pi$ certifying that $y = g^{2^T} \bmod N$. Following Wesolowski’s construction~\cite{wesolowski2020efficient}, a prime $q$ is derived from a hash of the inputs (Fiat-Shamir transformation~\cite{fiat1986prove}), and the worker computes $\pi = g^{\lfloor 2^T / q \rfloor} \bmod N$ with remainder $r = 2^T \bmod q$. The challenger then checks the relation $\pi^q \cdot g^r \equiv y \pmod N$, which holds if and only if the worker has correctly performed $T$ sequential squarings. 
This requires only multiple exponentiation operations, while the worker’s effort remains $O(T)$. The security of the scheme ensures that without knowledge of $\varphi(N)$, the adversary may not shortcut the sequential process.

To scale this construction for GPU-based measurement, we instantiate $C$ independent VDFs in parallel, where each instance $i$ uses a unique seed $(g_i, T_i) = (\mathsf{HashToQR}(\mathsf{sid}\,\|\,i),\, \mathsf{DeriveDelay}(\mathsf{sid}\,\|\,i))$. Each chain performs its own sequence of $T_i$ squarings, while the $C$ chains execute concurrently across GPU threads. This setup preserves per-instance sequentiality while fully utilizing the GPU’s parallel compute capacity.

For efficient procedure, the $C$ individual proofs $\{\pi_i\}$ are aggregated using random-looking coefficients $\{\alpha_i\}$ and a single prime $q$, both deterministically derived via a \emph{Fiat-Shamir challenge}~\cite{fiat1986prove}. Specifically, the worker and challenger independently compute
\begin{equation}
\begin{aligned}
(q, \alpha_1, \ldots, \alpha_C) \leftarrow
\mathsf{HashToPrimeAndScalars}\big(N,\{g_i, y_i, T_i\}_{i=1}^{C}, \mathsf{sid}\big),
\end{aligned}
\end{equation}

where the hash function acts as a public, non-interactive source of randomness, replacing the challenger's role in generating fresh challenges.
This transformation renders the framework fully non-interactive while preserving the soundness of the interactive variant under the random-oracle model~\cite{bellare1993random}.

Each instance produces its own output and proof as
$y_i = g_i^{2^{T_i}} \bmod N$,
$\pi_i = g_i^{\lfloor 2^{T_i}/q \rfloor} \bmod N$,
and $r_i = 2^{T_i} \bmod q$.
The challenger forms an aggregate proof
$\Pi = \prod_{i=1}^{C} \pi_i^{\alpha_i} \bmod N$
and checks the single batched relation
\begin{equation}
\Pi^{q} \cdot \Big(\prod_{i=1}^{C} g_i^{\alpha_i r_i}\Big)
\equiv
\prod_{i=1}^{C} y_i^{\alpha_i}
\pmod N.
\end{equation}

This greatly reduces challenger cost while maintaining the same security level as individual procedure.
The per-instance delay $T_i$ is chosen to align with the expected GPU execution time, where the nominal delay per instance is approximately $T_i \delta$, with $\delta$ denoting the average latency per modular squaring. Running $C$ VDF chains simultaneously enforces a measurable and verifiable computational load while guaranteeing that each instance remains inherently sequential. This \emph{multi-instance RSW-based VDF} construction thus provides a cryptographically sound, time-bound, and parallelizable framework for continuous GPU utilization measurement.

\section*{C. GEMM Puzzle-based Measurement}
\label{app:gemm_combined}

Our baseline VDF construction executes big-integer sequential squarings on CUDA cores. 
To better align with frontier AI workloads where tensor cores constitute the dominant execution pathway and account for over 90\% of the total compute, we introduce a complementary \emph{GEMM-based} measurement primitive whose core operation is a large dense matrix multiplication executed directly on tensor cores. 
Conceptually, this shifts the attested resource from general-purpose CUDA ALUs to the tensor-core GEMM pipeline.

Recent works shows that matrix multiplication can serve as an efficient Proof-of-Useful-Work primitive, allowing verification of $\mathsf{MatMul}(A,B)$ with negligible overhead on the prover side~\cite{komargodski2025proofs}. Cryptographic constructions based on \emph{trapdoored matrices} show how to verify linear-algebraic tasks such as matrix-vector and matrix-matrix multiplication, while preserving privacy and soundness via cryptographic trapdoors with near-quadratic complexity for verification~\cite{vaikuntanathan2025improving,braverman2025sublinear}. Inspired by these developments, we adopt a much simpler design tailored for GPU utilization measurement. We use \emph{vanilla dense GEMMs} compiled to tensor-core kernels and standard probabilistic verification by using Freivalds' algorithm~\cite{freivalds1977probabilistic}, without trapdoors or sophisticated encodings. This simplification yields a compute-based, tensor-core-centric utilization puzzle that integrates naturally with current GPU software stacks such as CUTLASS~\cite{cutlass}, cuBLAS~\cite{cublas}, and Triton~\cite{tillet2019triton}.

We consider a square GEMM of dimension $n$ (e.g., $n \approx 2^{13} \approx 10^4$), sized to saturate tensor cores and remain compute-bound.
Let $\mathbb{F}$ denote the floating-point domain with a fixed rounding mode, and let
$H:\{0,1\}^\ast \to \{0,1\}^\kappa$ be a cryptographic hash modeled as a random oracle.
A public difficulty parameter $d \in \{0,\dots,\kappa\}$ controls the expected number of puzzle attempts.
Given seed $\mathsf{sid}$, the challenger derives
$
\sigma_0 \leftarrow H(\mathsf{sid}).$
A deterministic matrix-derivation routine
$
\mathsf{DeriveMatrices}:\{0,1\}^\ast \to \mathbb{F}^{n\times n} \times \mathbb{F}^{n\times n}
$
expands $H(\sigma)$ into pseudorandom matrices $(A_\sigma,B_\sigma)$.

For attempt index $j \ge 0$, the puzzle iterates
\[
\sigma_{j+1} = H(\sigma_j),\qquad
(A_j,B_j)=\mathsf{DeriveMatrices}(\sigma_j),\qquad
C_j = A_jB_j,
\]
where $C_j$ is computed on GPU tensor cores.
The worker forms
$h_j = H(\mathsf{sid}\,\|\,\sigma_j\,\|\,\|\,C_j)$
and succeeds if
\begin{equation}
h_j < 2^{\kappa-d}.
\label{eq:gemm-threshold-final}
\end{equation}
Attempts succeed with probability $2^{-d}$, giving a geometric distribution with expectation $2^d$.
A valid proof consists of $(idx=j^\star,\ C_{j^\star})$.

To validate a proof $(j^\star,C_{j^\star})$, the challenger:
\begin{enumerate}
  \item Recomputes $\sigma_{j^\star}$ via the hash chain.
  \item Computes $(A_{j^\star},B_{j^\star})=\mathsf{DeriveMatrices}(\sigma_{j^\star})$.
  \item Re-evaluates
$  h_{j^\star}' =
  H(\mathsf{sid}\,\|\,\sigma_{j^\star}\,\|\,C_{j^\star})$
  and checks Equation~\eqref{eq:gemm-threshold-final}.
  \item Verifies that $C_{j^\star}$ is a correct (or sufficiently accurate) product using Freivalds' algorithm~\cite{freivalds1977probabilistic}.
\end{enumerate}

The challenger’s cost is
$T_{\mathsf{verify}}(n,k,j^\star)
  = O(k n^2 + j^\star \cdot \mathsf{HashCost}),$
dominated by the $k n^2$ multiplications in Freivalds’ checks.
The expected worker cost is
$\mathbb{E}[T_{\mathsf{worker}}] = \Theta(2^d n^3)$
floating-point operations.
Selecting $n$ to match hidden dimensions in large models (e.g., $8\text{k}\text{--}16\text{k}$)~\cite{team2025longcat}
ensures the puzzle remains strongly compute-bound while detection stays lightweight.

\noindent\textbf{Freivalds Verification of GEMM. }
Freivalds’ algorithm checks the equality
$C_{j^\star} = A_{j^\star}B_{j^\star},$
without recomputing the full GEMM.
Each round samples a random vector $r$, compares
$A_{j^\star}(B_{j^\star} r)$ with $C_{j^\star} r$, and rejects on mismatch. 
After $k$ successful iterations, the probability of accepting an incorrect product is at most $2^{-k}$.
A final hash-threshold check enforces consistency with the puzzle.

We note that Freivalds' algorithm is perfectly accurate over a finite field $\mathbb{Z}_p$. 
However, when implemented using floating-point arithmetic, discrepancies such as $y^{(\ell)} \neq z^{(\ell)}$ may arise due to small rounding errors, even when the GPU has performed the computation ``correctly.''
A common remedy in such systems is to define the GEMM operation over a large prime field using modular arithmetic, ensuring that the equality $y = z$ holds exactly. 
If floating-point computations must be used, the verification should instead check whether $|y - z| < \epsilon$ for a small tolerance $\epsilon$. 

\begin{algorithm}[t]
\small
\caption{Freivalds-Based Verification of Matrix Product $C_{j^\star} = A_{j^\star}B_{j^\star}$}
\label{alg:freivalds}
\DontPrintSemicolon
\KwIn{
  Matrices $A_{j^\star}, B_{j^\star}, C_{j^\star} \in \mathbb{F}^{n \times n}$;\\
  security parameter $k$
}
\KwOut{\textbf{ACCEPT} or \textbf{REJECT}}

\For{$\ell \gets 1$ \KwTo $k$}{
    Sample $r^{(\ell)} \sim \{0,1\}^n$\;
    Compute $x^{(\ell)} \gets B_{j^\star}\, r^{(\ell)}$\;
    Compute $y^{(\ell)} \gets A_{j^\star}\, x^{(\ell)}$\;
    Compute $z^{(\ell)} \gets C_{j^\star}\, r^{(\ell)}$\;
    \lIf{$y^{(\ell)} \neq z^{(\ell)}$}{
        \textbf{REJECT}; \textbf{break}
    }
}
\If{all $k$ iterations passed}{
    Verify hash-threshold condition\;
    \lIf{\text{condition holds}}{\textbf{ACCEPT}}
    \lElse{\textbf{REJECT}}
}
\end{algorithm}

\section*{D. Telemetry Observability Considerations}
\label{app:considerations_full}

\noindent\textbf{Timing Resolution and Statistical Behavior}
The ability to observe deviations in resource utilization depends on the timing characteristics of each measurement class.  
In PoW and GEMM-based workloads, solve-time distributions follow exponential patterns, with rate $\Lambda = R p M$, where $M$ is the number of active threads.  
Deviations from the expected solve rate, whether faster or slower, may indicate resource contention, offloading, or co-location interference.  
In contrast, the VDF-based timing model is nearly deterministic, reflecting its strictly sequential nature and offering more stable temporal inference.

We model each class's resolution behavior using standard probabilistic tools.  
Given a target confidence factor $P_{\mathrm{honest}}$ (e.g., $0.9$), one can derive a minimum test count $N$ and timing threshold $\tau$ such that deviation is statistically distinguishable from expected solve times.  
We describe a family of iterative estimators and detection curves for each measurement class. 

\noindent\textbf{Observability Under Outsourcing}
Although these measurements are not designed for formal attestation, they produce latency patterns that are sensitive to compute locality.  
In PoW workloads, memory-hard primitives such as \texttt{Argon2id} induce high bandwidth and random access demand, which are difficult to emulate efficiently on CPUs or ASICs.  
GEMM-based puzzles require high-throughput tensor operations that scalar architectures cannot reproduce without substantial timing shifts.  
VDF workloads, while sequential, scale poorly on CPU clusters due to limited parallel throughput and inter-node communication costs.

These effects are reflected in timing distributions: outsourcing often results in detectable round-time inflation or increased variability, offering potential indicators of execution environment.

\noindent\textbf{Energy Overhead and Adaptability}
Each measurement class introduces different overhead characteristics.  
PoW and GEMM incur sustained compute and memory pressure, and thus higher power draw at higher sampling rates.  
VDF evaluations, while still compute-intensive, involve bounded sequential workloads with fewer memory iterations, reducing average energy footprint.  
By modulating puzzle issuance rate and difficulty, the challenger can adjust the trade-off between measurement sensitivity and energy cost, informing future deployment strategies.

\noindent\textbf{Timing Leakage and Privacy Surfaces}
While the constructions transmit no identifying information, response timing may correlate with background workloads.  
Solve-time variation induced by co-resident models (e.g., large LLMs) could expose coarse-grained behavioral patterns, depending on scheduling and load.  
These effects are not unique to our scheme and arise in any time-sensitive measurement framework lacking hardware isolation.  
Padding, randomized delays, or noise injection may mitigate timing-based leakage.

\section*{E. Additional Results}
\label{app:additional_results}

This appendix reports supplementary experimental results that further characterize the behavior of the proposed measurement mechanisms under alternative configurations, extended parameter sweeps, and additional contention scenarios. 

\begin{figure}[t]
\centering
\includegraphics[width=0.8\columnwidth]{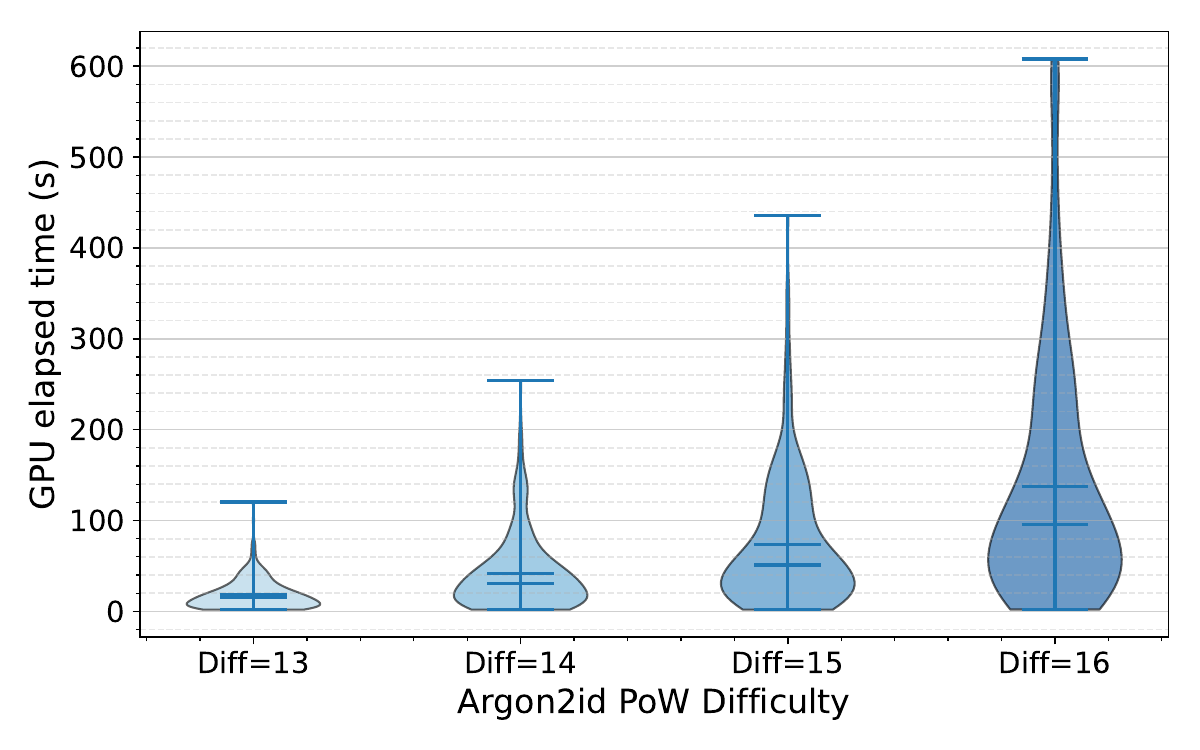}
\caption{Violin plot of resolution time of PoW puzzles for $N=300$ trials across four difficulty levels.}
\label{fig:gpu_times_violin_by_diff}
\end{figure}

\noindent\textbf{PoW Difficulty Scaling. }
Figure~\ref{fig:gpu_times_violin_by_diff} provides an alternative visualization of PoW resolution times across multiple difficulty levels using violin plots. This representation highlights both the central tendency and variance of solve times, reinforcing the exponential scaling behavior observed in the histogram-based analysis.

\begin{figure}[t]
\centering
\includegraphics[width=0.8\columnwidth]{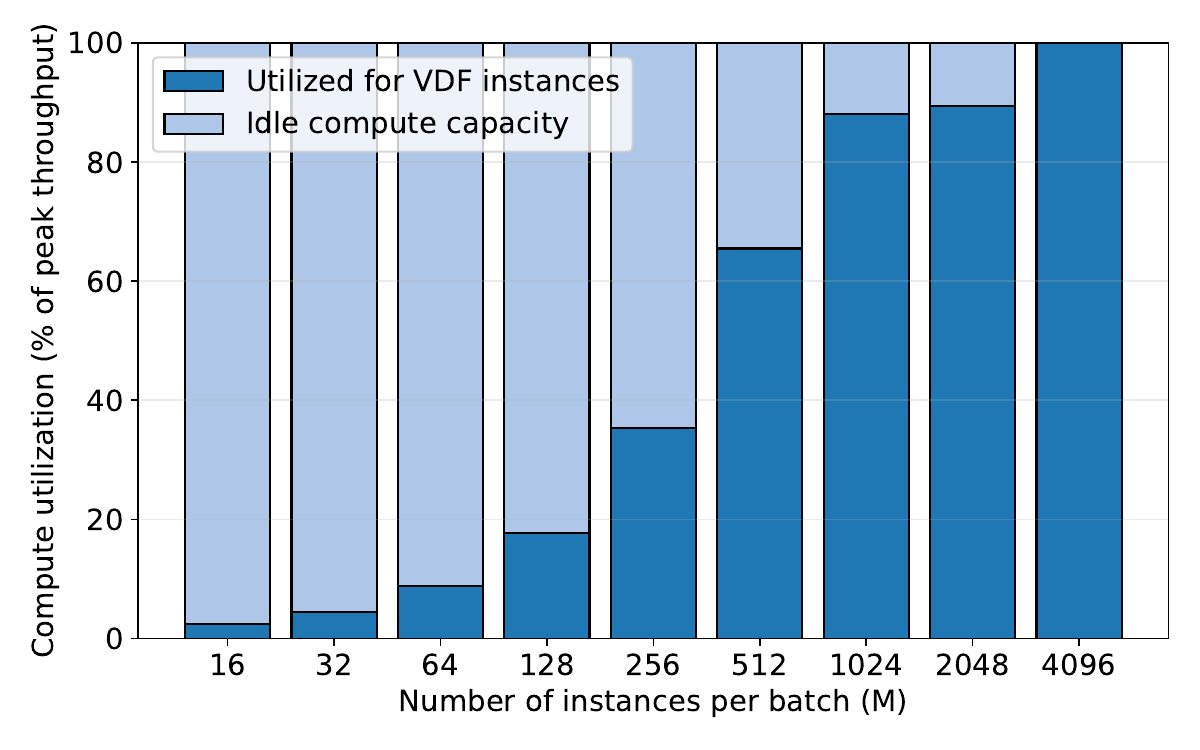}
\caption{Empirical GPU utilization as a function of the number of parallel VDF instances for fixed puzzle difficulty.}
\label{fig:gpu_compute_utilization_vs_M}
\end{figure}

\noindent\textbf{VDF Instance Scaling and GPU Saturation}
We further analyze how increasing the number of parallel VDF instances affects GPU utilization. Figure~\ref{fig:gpu_compute_utilization_vs_M} shows the normalized compute utilization as a function of the number of concurrent VDF instances for a fixed difficulty parameter. The results indicate rapid saturation of CUDA cores, with diminishing returns beyond a certain instance count.

\begin{figure}[t]
\centering
\includegraphics[width=0.8\columnwidth]{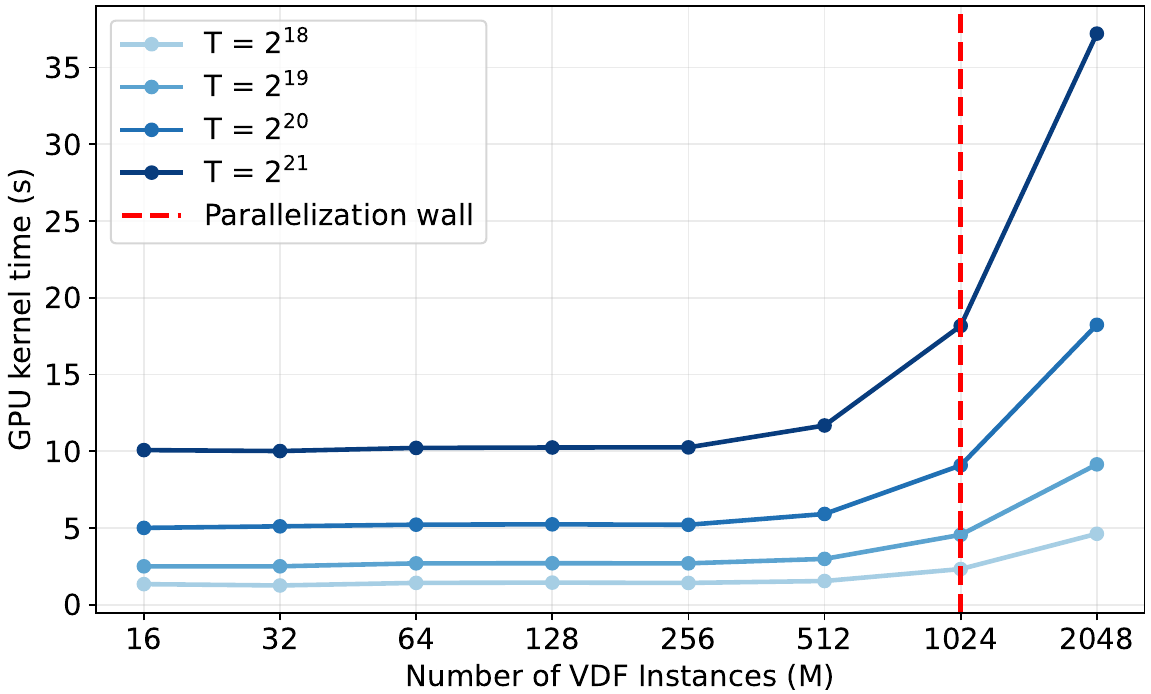}
\caption{Kernel runtime scaling across varying numbers of VDF instances and difficulty parameters.}
\label{fig:kernel_vs_M_multiT}
\end{figure}

Figure~\ref{fig:kernel_vs_M_multiT} further illustrates kernel runtime scaling across different instance counts and difficulty parameters, confirming a near-linear increase once the GPU reaches full occupancy.

\begin{figure}[t]
\centering
\includegraphics[width=0.8\columnwidth]{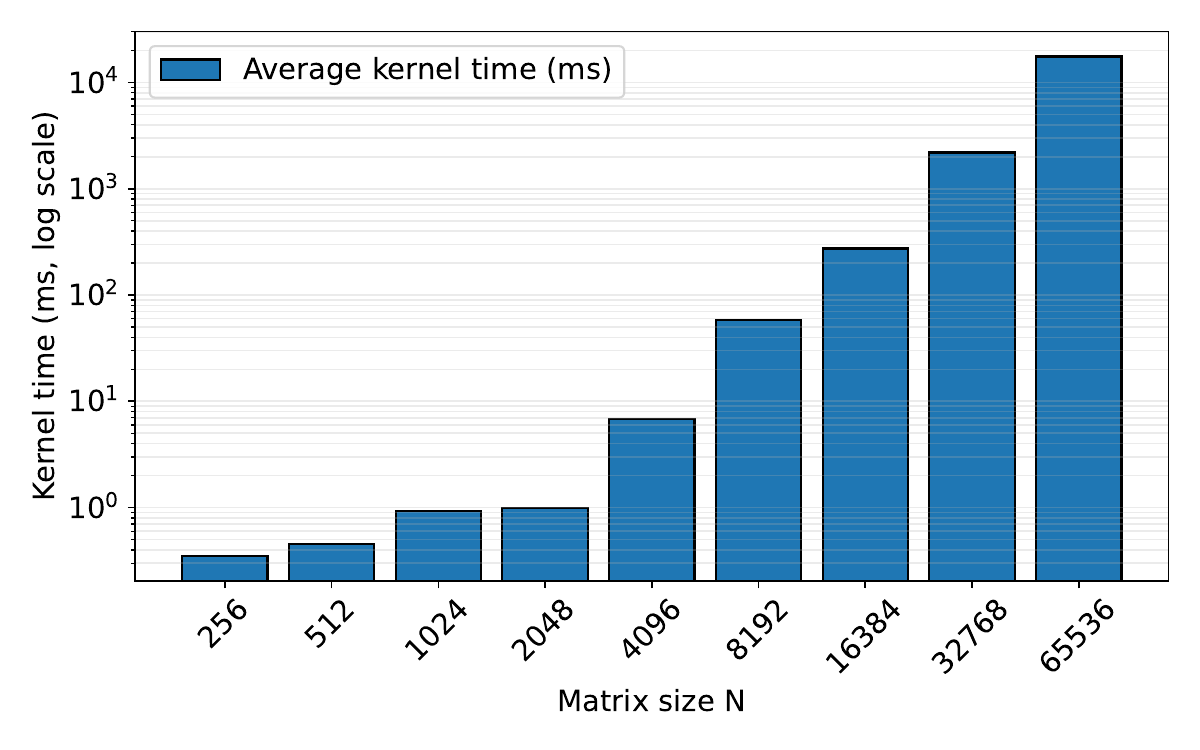}
\caption{Kernel runtime for GEMM-based PoW puzzles as a function of matrix dimension $N$ on H100.}
\label{fig:gemm_kernel_time_vs_N}
\end{figure}

\noindent\textbf{GEMM Dimension Scaling}
In addition to difficulty-based scaling, we examine how the matrix dimension affects the execution time of GEMM-based PoW puzzles. Figure~\ref{fig:gemm_kernel_time_vs_N} shows the kernel runtime for increasing matrix sizes on an H100 GPU using CUTLASS-based implementations. The results confirm that runtime grows predictably with matrix dimension, enabling fine-grained tuning of compute cost.

\end{document}